\documentclass[10pt,journal,compsoc]{IEEEtran}
\usepackage{tcolorbox}

\usepackage{parcolumns}
\usepackage{amsmath,amssymb,amsfonts}
\usepackage{booktabs,makecell,tabularx}
\usepackage{algorithmic}
\usepackage{graphicx}
\usepackage{textcomp}
\usepackage{xcolor}
\usepackage{lipsum}
\usepackage{framed}
\usepackage{color}
\usepackage{soul}
\usepackage[export]{adjustbox}
\usepackage{floatrow}
\usepackage{hyperref}
\usepackage{algorithm}
\usepackage{subfiles}
\usepackage{verbatim}
\usepackage{listing}
\usepackage{listings}   
\usepackage{enumitem}
\usepackage{tikz}
\usepackage{multirow}
\usepackage{multicol} % Allow spanning cells in tables
\usepackage{pifont} % for the \ding{} Zapf dingbats font

\newcommand{\xmark}{\ding{55}}

\newcommand{\subsubsubsection}[1]{\paragraph{\textit{#1}}}

\ifCLASSOPTIONcompsoc
  \usepackage[nocompress]{cite}
\else
  \usepackage{cite}
\fi
\ifCLASSINFOpdf
\else
\fi
\newcommand{\ali}[1]{\textcolor{blue}{{\it}}}
\newcommand{\mehdi}[1]{\textcolor{red}{{\it}}}
\newcommand{\code}[1]{\textcolor{codeColor}{\textsf{#1}}}

\newcommand{\figref}[1]{Figure~\ref{#1}}
\newcommand{\secref}[1]{Section~\ref{#1}}

\newcommand{\equationref}[1]{Equation~\eqref{#1}}
\newcommand{\tableref}[1]{Table~\ref{#1}}
\definecolor{codeColor}{RGB}{139,26,26}

\renewcommand{\hl}[1]{#1}
\newcommand{\citeme}[1]{%
  \begingroup
  \definecolor{hlcolor}{RGB}{255, 226, 176}\sethlcolor{hlcolor}%
  \textcolor{darkgray}{\hl{\textbf{CITE}}}%
  \endgroup
}

\newcolumntype{P}[1]{>{\centering\arraybackslash}p{#1}}

\newcommand\shorteq{%
 \rule[.4ex]{4pt}{0.4pt}\llap{\rule[.7ex]{4pt}{0.4pt}}}

\begin{document}
\title{\textsc{IPSynth}: Interprocedural Program Synthesis for Software Security Implementation}

%IPSYNTH: An Inter-procedural Software Synthesis Approach for Architectural Tactic Implementation

\author{Ali~Shokri,
        Ibrahim~Jameel~Mujhid,
        and~Mehdi~Mirakhorli% <-this % stops a space
\IEEEcompsocitemizethanks{\IEEEcompsocthanksitem All authors are with the Department of Software Engineering, Rochester Institute of Technology, Rochester,
NY, USA. (email: as8308@rit.edu; ijmvse@rit.edu; mxmvse@rit.edu)}% <-this % stops an unwanted space
\thanks{}}

\IEEEtitleabstractindextext{%
\begin{abstract}
To implement important quality attributes of software such as architectural security tactics, developers incorporate API of software frameworks, as building blocks, to avoid re-inventing the wheel and improve their productivity. However, this is a challenging and error-prone task, especially for novice programmers. 
Despite the advances in the field of API-based program synthesis, the state-of-the-art suffers from a twofold shortcoming when it comes to architectural tactic implementation tasks. First, the specification of the desired tactic must be explicitly expressed, which is out of the knowledge of such programmers. Second, these approaches synthesize a block of code and leave the task of breaking it down into smaller pieces, adding each piece to the proper location in the code, and establishing correct dependencies between each piece and its surrounding environment as well as the other pieces, to the programmer.

To mitigate these challenges, we introduce IPSynth, a novel inter-procedural program synthesis approach that automatically learns the specification of the tactic, synthesizes the tactic as inter-related code snippets, and adds them to an existing code base. We extend our first-place award-winning extended abstract recognized at the 36th IEEE/ACM International Conference on Automated Software Engineering (ASE'21) research competition track. In this paper, we provide the details of the approach, present the results of the experimental evaluation of IPSynth, and analyses and insights for a more comprehensive exploration of the research topic. Moreover, we compare the results of our approach to one of the most powerful code generator tools, ChatGPT. Our results show that our approach can accurately locate corresponding spots in the program, synthesize needed code snippets, add them to the program, and outperform ChatGPT in inter-procedural tactic synthesis tasks.

\end{abstract}

}

\maketitle

\IEEEdisplaynontitleabstractindextext

% \fi
\IEEEpeerreviewmaketitle

\IEEEraisesectionheading{\section{Introduction}\label{sec:Introduction}}
\IEEEPARstart{S}oftware architectural tactics are re-usable design strategies that aim to preserve important quality attributes of a software system, including security, interoperability, availability, performance, and scalability \cite{mirakhorli2015modifications}. 
To satisfy these concerns, software architects carefully compare and choose proper architectural tactics to be implemented by software developers in the code~\cite{Bass, ICSE2012}. %\cite{Bass, ICSE2012}. 
In the subsequent development phase, these architectural choices must be implemented completely and correctly in order to avoid any drifts from the envisioned design~\cite{van2005design}.
% Bass et al.~\cite{Bass} provide a comprehensive list of such tactics and define security tactics as reusable solutions to satisfy security quality attributes regarding resisting attacks (e.g., tactic ``Authenticate Actors''), detecting attacks (e.g., tactic ``Detect Intrusion''), reacting to attacks (e.g., tactic ``Revoke Access''), and recovering from attacks (e.g., tactic ``Audit''). 

% The importance of the correct tactic implementation has been highlighted by several earlier studies~\cite{TacticLatentTopics, mirakhorli2015modifications}. These studies found that even when suitable tactics are chosen upfront, developers\textemdash especially less experienced ones\textemdash often struggle when implementing these tactics in the code~\cite{rehman2018roles} and implement them incorrectly, causing software bugs and flaws~\cite{SAINICSA}. 

Previous studies~\cite{TacticLatentTopics, mirakhorli2015modifications} emphasize the significance of correct tactic implementation. In practice, developers\textemdash especially less experienced ones\textemdash face challenges in correctly implementing chosen tactics\cite{rehman2018roles}, resulting in software bugs and flaws~\cite{SAINICSA}.
Insufficient experience, limited knowledge of architectural tactics, and unfamiliarity with the underlying framework and the code base are primary factors contributing to an incorrect implementation of these tactics \cite{shokri2021_icsa}. 
%
%As a case in point, the study by Santos et.al.~\cite{santos2017understanding} demonstrates that adding \textit{\textbf{authentication}} and \textit{\textbf{authorization}} security tactics to a program is a nontrivial and error-prone task for such programmers. 
In practice, these tactics are mostly implemented by incorporating APIs of third-party libraries \cite{cervantes2012principled}. For example, Java Authentication and Authorization Security Services (JAAS) \cite{JAAS} is a popular Java-based framework that provides APIs for adding such tactics to a program.

Program synthesis has been promising in providing support for programmers by automatically constructing a program in an underlying programming language, based on a given specification of the to-be-synthesized program \cite{gulwani2017program}. 
%Program developers can benefit from the synthesized code, either by directly reusing it in their programming tasks or by learning from it.
%Component-based program synthesis \cite{jha2010oracle}, which aims to create a code snippet only from a list of given components (e.g., APIs), could be an answer to this need. 
In the past recent years, there have been works introduced by researchers that specifically focus on synthesizing a program from a given set of API calls \cite{feng2017component, yang2018edsynth, shi2019frangel, guo2019program, liu2020prosy, liu2020much}. However, despite the advances in this field, the current state-of-the-art API-based program synthesis approaches suffer from a two-fold shortcoming when it comes to architectural tactic implementation tasks. First, the specification of the desired tactic has to be explicitly expressed by the user, which comes with complexity~\cite{rehman2018roles} and is out of the knowledge of non-expert programmers~\cite{van2005design}. Second, these approaches synthesize a block of code and leave the task of breaking down the synthesized code into smaller pieces, adding each piece to its proper location in the code, and establishing correct dependencies between each piece and its surrounding environment as well as the other pieces, to the programmer. This is obviously a non-trivial task, even for more savvy programmers.

To mitigate these challenges, in this paper, we introduce \textsc{IPSynth}, a novel inter-procedural program synthesis approach that automatically learns the specification of the to-be-synthesized tactic, synthesizes the tactic as inter-related code snippets, adds each piece to its corresponding location in the given code base, and establishes correct dependencies (i.e., control- and data-dependencies) between each piece and its surrounding environment. 
We particularly enhance and extend our prior extended abstract, titled "A program synthesis approach for adding architectural tactics to an existing code base" \cite{shokri2021program} which was awarded the first-place prize in the research competition track at ASE'21. Building on the positive feedback and encouragement received, this journal paper provides a more in-depth and rigorous investigation of the research subject.
%The synthesizer follows the concept of \textit{correct-by-construction}, meaning that it makes sure that the synthesis process does not generate an incorrect (semantically and syntactically) tactic. 

The purpose of \textsc{IPSynth} is: \textit{given a compilable program, i.e., a syntactically correct program, it is able to automatically synthesize and add architectural tactics to that program such that the final program is syntactically and semantically (w.r.t. tactic implementation) correct.}

\subsection{Contributions}
In summary, the contributions of this paper are as follows.
\begin{itemize}
    \item A novel inter-procedural program synthesis approach  called \textsc{IPSynth} for adding architectural tactics to a program. This approach follows API-based program synthesis in which the specification is automatically inferred from a pre-learned API usage specification model as well as the context of the under-development program.
    \item A dataset of test programs to be used by other researchers for inter-procedural program synthesis tasks. The dataset consists of $20$ architectural tactic implementation tasks and is publically available from its anonymized repository at \url{https://anonymous.4open.science/r/Anonymous-82DE}.   
    \item An experimental study of the approach to investigate its accuracy and effectiveness. This study has been conducted in two different levels of granularity, (i) component level where we assess the functionality of each of the components of the approach separately, and (ii) the entire approach, where we evaluate the accuracy of the synthesized code and compare it against the output of one of the most powerful tools that is widely used by programmers, ChatGPT \cite{OpenAI2023GPT4TR}. The experiment and the results are provided in \secref{sec:IPSynth-experimentalStudy}. %As will be presented  results show, \textsc{IPSynth} is able to accurately implement tactics and outperforms related work such as ChatGPT. %\ali{ We also provide a case study in which we investigate the performance of our approach in a real-life program synthesis scenario.} 
\end{itemize}

The rest of the paper is structured as follows. \secref{sec:MotivatingExample} motivates our work through a tactic synthesis example. The approach is detailed in \secref{sec:IPPS}. We provide an experimental study of the approach in \secref{sec:IPSynth-experimentalStudy}. \secref{sec:Limitation} discusses the limitation of the approach sheds light on the future direction of the work. Related works are studied in \secref{sec:relatedWork}. Finally, \secref{sec:Conclusion} concludes the paper.

\section{Motivating Example}
\label{sec:MotivatingExample}
In order to better motivate our problem, in this section, we provide an example of an architectural tactic synthesis task. We then, use this example throughout the paper to provide a better walk-through of the technical parts of the approach. 
\begin{figure}
    \centering
    \includegraphics[width=\textwidth]{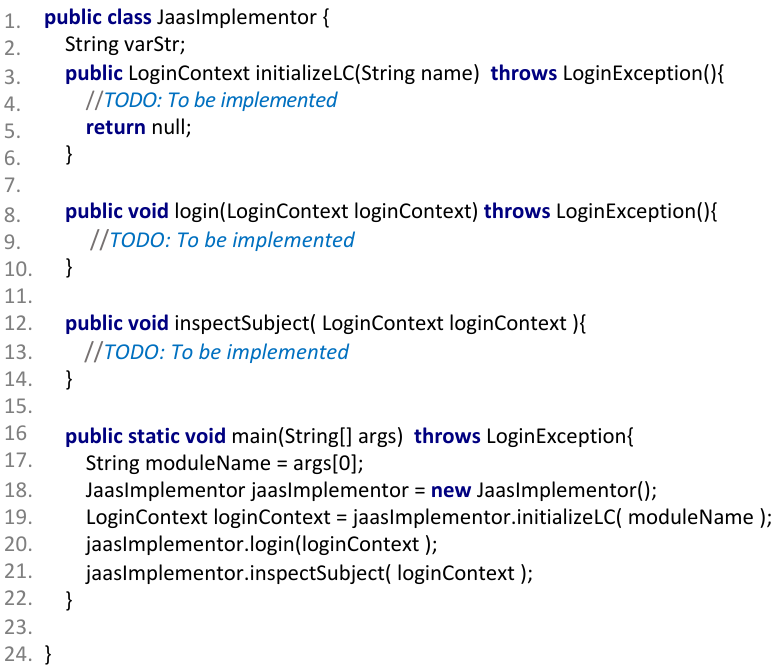}
    \caption{Before synthesizing the \textit{authentication} tactic}
    \label{fig:MotivatingExample_UDP}
\end{figure}
\begin{figure}
    \centering
    \includegraphics[width=\textwidth]{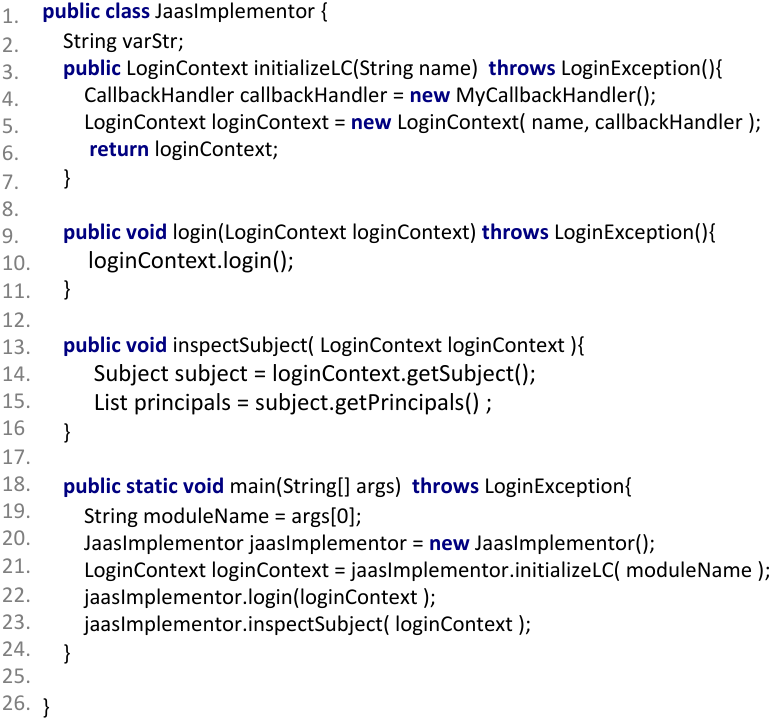}
    \caption{After synthesizing the \textit{authentication} tactic}
    \label{fig:MotivatingExample_UDP_Synthesized}
\end{figure}
\figref{fig:MotivatingExample_UDP} shows an under-development program in which the programmer aims to use \textit{Java Authentication and Authorization Services (JAAS)}, a popular open-source java-based framework, to implement \textit{authentication} security tactic and restrict non-user access to unwanted resources. This would not be a trivial task, especially for novice programmers. The entry point of this program is the \code{main} method (line 16), which retrieves the name of the authentication module from the \code{args} argument (line 17) and creates an instance of \code{JaasImplementor} class (line 18). Then, in a sequence of method calls, it is supposed to instantiate needed objects for the authentication process (line 19), perform the login process (line 20), and finally, inspect the authenticated subject (line 21). This is a routine process one needs to follow for an authentication task. While these steps seem limited, the challenge comes from \textbf{(i)} creating pieces of code snippets constructed from correct APIs to be used in different parts of the program, \textbf{(ii)} identifying candidate locations in the program in which each piece can be added to, and \textbf{(iii)} correctly connecting each piece to its surrounding context, as well as to the other pieces.
\figref{fig:MotivatingExample_UDP_Synthesized} shows a correct implementation of authentication tactic using the JAAS framework for the program shown in \figref{fig:MotivatingExample_UDP}. To start the authentication process, an object of \code{LoginContext} should be instantiated (line 5). \code{LoginContext} is the keystone part of an authentication process implemented by the JAAS framework. However, the constructor of \code{LoginContext} requires an object of \code{CallbackHandler} as an input argument. Hence, an object of \code{CallbackHandler} is created at line 4. A \code{CallbackHandler} handles the communication with the user, including collecting the user's \textit{username} and \textit{password}. Once the \code{LoginContext} is created and returned (line 6), it can be used to perform the actual logging-in process. This is implemented by calling the \code{login()} method of the previously created \code{LoginContext} in a separate method (line 10). Finally, to inspect the output of the authentication process, this program retrieves the \code{subject} object (line 14) that is populated during the login process. A \code{Subject} encompasses all the information around an authenticated user. In case of successful authentication, the retrieved \code{subject} has the user's \code{principals} (line 15), which can be corresponding \textit{roles} of the user in the specified context. Otherwise, the returned set of \code{principals} would be empty. There are some more details around the JAAS framework which we have overlooked in this example for the sake of simplicity.  

As the example demonstrates, the task of program synthesis in architectural tactic implementation is an inter-procedural task. It means that APIs might be used in different methods of different classes in a program, yet they need to interact with each other through method calls. Moreover, to guarantee the semantic correctness of the synthesized program, the synthesizer needs to have a good knowledge of correct control- and data dependencies between the used APIs in the program for implementing a specific tactic. Also, it is crucial to analyze the under-development program and spot the correct locations in the code to which each piece of the tactic should be added. Finally, for each spot, the corresponding code snippet should be synthesized such that the correct data- and control dependencies between that code snippet and its surrounding environment, as well as the other synthesized code snippets, are established.

The state-of-the-art approaches are not designed and equipped for such a task. The approach introduced in this paper aims to address the mentioned needs. \figref{fig:MotivatingExample_UDP_Synthesized} represents the outcome of our approach for synthesizing the authentication tactic for the program shown in \figref{fig:MotivatingExample_UDP}.

\section{Inter-Procedural Program Synthesis}
\label{sec:IPPS}

\begin{figure*}[ht]
    \centering
    \includegraphics[width=\textwidth]{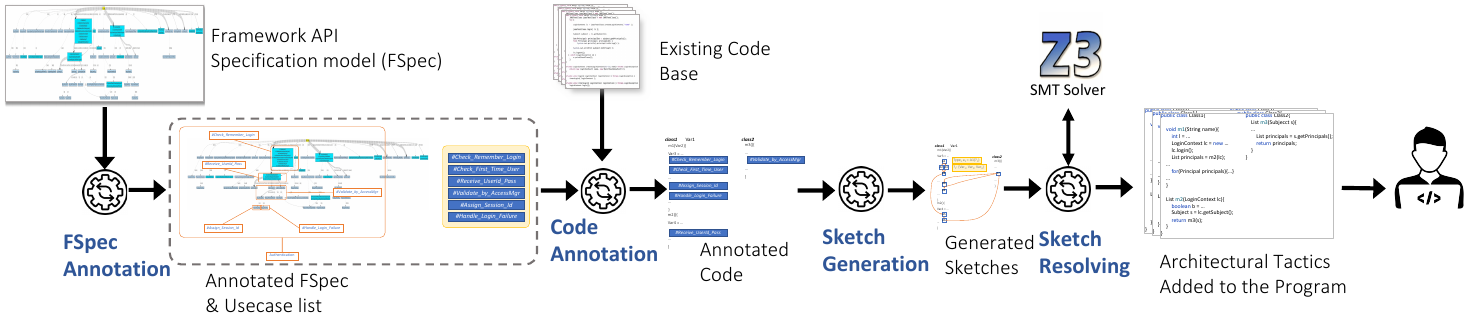}
        \centering
    \caption{An overview of our \textsc{IPSynth} approach}
    \label{fig:ipps}
\end{figure*}

%As discussed before in \secref{sec:Introduction}, the current state-of-the-art program synthesis approaches cannot \hl{properly} handle inter-procedural program synthesis tasks, such as architectural tactic implementation. 
To address the gaps mentioned in the current state-of-the-art approaches for synthesizing architectural tactics in a program, we introduce our \textsc{\textbf{I}nter-procedural \textbf{P}rogram \textbf{Synth}esis} (\textbf{\textsc{IPSynth}}) approach. \figref{fig:ipps} provides an overview of \textsc{IPSynth} which consists of four main steps. Before providing a birds-eye view of the steps, we briefly provide background about our prior work, \textsc{ArCode} \cite{shokri2021_icsa, shokri2021_icpc}, that produces specification models of APIs that programmers use in their tactic implementation tasks. We use this model during the first step of \textsc{IPSynth} (\secref{sec:IPPS_FSpecAnnotation}).

%In the first step (\secref{sec:IPPS_FSpecAnnotation}), we learn and create clusters of API calls that contribute to the implementation of different parts of a tactic. To scope our contributions in this paper, we chose to use our previous work, ArCode \cite{shokri2021_icsa, shokri2021_icpc}, for creating  graph-based API specification models for APIs that programmers use in tactic implementation tasks. Following, we provide a brief background of ArCode, however, the full details can be found in its papers.

\begin{figure}[h]
    \centering
    \includegraphics[width=.7\linewidth]{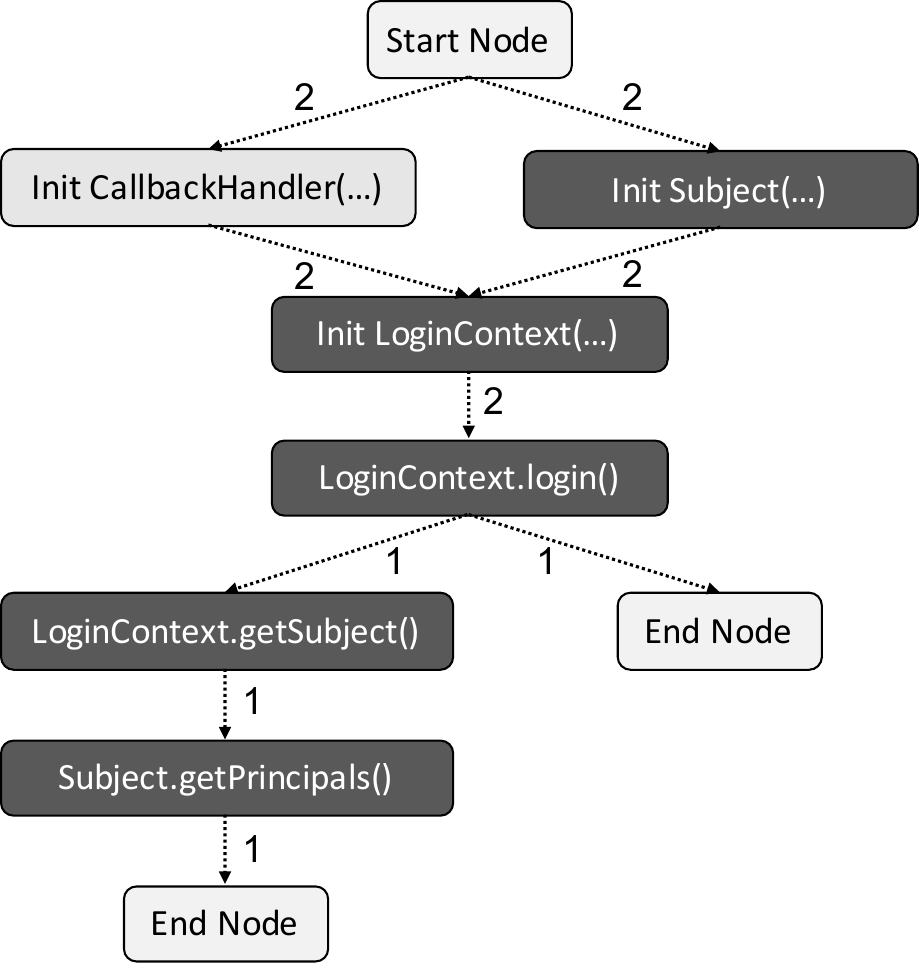}
        \centering
    \caption{An FSpec constructed by \textsc{ArCode} that demonstrates two correct ways of incorporating JAAS API for implementing the authentication tactic.}
    \label{fig:FSpec}
\end{figure}

\textsc{\textbf{ArCode}} is a learning-by-example approach designed for learning API specifications related to architectural tactic implementation tasks. ArCode is able to generate API specification models (FSpec) for the frameworks that developers incorporate to implement tactics (e.g., Java Authentication and Authorization Services - JAAS). \textit{FSpec} is a directed graph that the nodes are API calls and the edges are dependencies between API calls (e.g., control- and data-dependencies) that one should consider in order to \textit{correctly} implement a tactic using those API calls. An FSpec represents the correct ways of using API calls in order to \textbf{correctly} implement a tactic. \figref{fig:FSpec} demonstrates an example of an FSpec generated by \textsc{ArCode} that represents two ways to correctly use API calls provided by the JAAS framework for implementing the authentication tactic. Each correct API usage starts with the \textit{start} node and ends in an \textit{end} node. The numbers on the edges of this model represent the frequency of times that \textsc{ArCode} observed the same tactic in different programs in its training set.

In the first step of \textsc{IPSynth} (\secref{sec:IPPS_FSpecAnnotation}), we perform clustering on API calls inside the FSpec model such that each cluster corresponds to a tactic-related sub-task \textemdash conceptual sub-tasks in the realization of a tactic. For example, in the case of the \textit{authentication} tactic implementation by incorporating the JAAS framework, there will be API clusters related to \textit{object instantiation}, \textit{logging in}, and \textit{subject inspection}. We also annotate the clusters with meaningful labels, indicating which part of the FSpec model implements what sub-task of the tactic implementation. Next, we find a mapping between the annotated clusters in the FSpec and the correct locations in the code that tactic pieces should be synthesized and added to (\secref{sec:IPPS_CodeAnnotation}). In the third step (\secref{sec:IPPS_SketchGeneration}), sketches of to-be-synthesized pieces of the tactic will be generated. A \textit{sketch} is a skeleton of a code snippet that consists of some unimplemented parts, i.e. \textit{holes}. Lastly, by resolving the holes in sketches (\secref{sec:IPPS_SketchResolving}), we generate an actual tactic implementation from the previously created code structure. 
This approach follows the concept of \textit{correct by construction}, meaning that we make sure that the synthesis process does not generate an incorrect (semantically and syntactically) tactic. The details of the mentioned steps are provided as follows. 

%, namely FSpec annotation (\secref{sec:IPPS_FSpecAnnotation}), code annotation (\secref{sec:IPPS_CodeAnnotation}), sketch generation (\secref{sec:IPPS_SketchGeneration}), sketch resolving (\secref{sec:IPPS_SketchResolving}), and verification (\secref{sec:IPPS_Verification}). While in the first two steps, IPSynth automatically extracts the specification of the to-be-synthesized tactic, it synthesizes the tactic through the last three steps. In the followings, we provide details on each step. 

\subsection{Step \#1 - FSpec Annotation}
\label{sec:IPPS_FSpecAnnotation}
Since we follow the API-based program synthesis approach, we need to learn how to correctly use the API of a framework for correctly implementing a tactic in a program. In that regard, we leverage \textsc{ArCode} \cite{shokri2021_icsa, shokri2021_icpc}, our prior work which creates the \textit{framework API specification model (FSpec)}, a probabilistic model that represents only the correct ways of incorporating API of that framework in a program to implement a tactic. 
To better guide our program synthesizer, we cluster API calls inside the FSpec such that all APIs inside a cluster contribute to the implementation of a (sub)task. Then, we \textbf{annotate} (i.e., label) each cluster with a meaningful name that represents the task that APIs inside the cluster contribute to. An annotation can be a simple or a compound name. For instance, in the code snippet provided in \figref{fig:MotivatingExample_UDP_Synthesized}, creating instances of \code{CallbackHandler} and \code{LoginContext} (line 4-5) would be considered as the task of \textit{initializing} objects needed for the authentication process. Therefore, one would expect to see these two API calls in the same cluster inside the FSpec with an annotation such as \textbf{\textit{\#Initialization}}. In general, it would be quite possible for API calls of different clusters to collaborate and contribute to more coarse-grained tasks. Therefore, we will end up in a hierarchical clustering/annotation structure. For example, one would expect to see a couple of API calls in a cluster annotated as \textit{\#Authentication}, while there are  more fine-grained clusters annotated as \textit{\#Initialization}, \textit{\#Logging\_In}, and \textit{\#Subject\_Inspection} inside the \textit{\#Authentication}. Ideally, generated code for each fine-grained cluster will be put in a separate method, yet, methods will be collaborating to implement the entire tactic.
%Throughout this dissertation, we might use terms ``FSpec annotations'' and ``tactic pieces'' interchangeably.

To automatically generate clusters and their corresponding annotations, we follow an iterative process that starts with each API call inside the FSpec. The process is explained as follows.
\begin{enumerate}
    \item We extend the \textsc{ArCode} approach such that during the FSpec creation, we generate an annotation for each API call inside the FSpec. The annotation of an API call would represent the finest-grained (sub)task that the API call contributes to. For this purpose, let's say that API $A_1$ in the FSpec was found from method $m$ of class $c$ of program $p$ during the FSpec creating process. We aim to generate a short summary of the task that method $m$ contributes to and annotate API $A_1$ inside the FSpec with that summary. 
    
    To achieve this goal, we leverage state-of-the-art \textit{method naming suggestion} approaches that find meaningful names for a given method. In particular, we use Code2Vec \cite{alon2019code2vec} which is an approach that given an implementation of a method (i.e., the body of the method), finds semantically similar methods in its code repository and returns a ranked list of suggested methods names for the given method implementation. In other words, based on a given method implementation, it returns a ranked list of meaningful name suggestions that aim to describe the semantics of the given implementation. We use this technique to annotate each API call inside the FSpec model.
    For example, if \textsc{ArCode} finds \code{CallbackHandler} initialization in method \code{m} of class \code{c} of program \code{p} in a code repository that is used for creating the FSpec, we pass the body of method \code{m} to Code2Vec and receive a ranked list of corresponding names. Then, we use the 1st ranked name as the annotation for \code{CallbackHandler} initialization API inside the FSpec. Please note that since an API could be found from multiple methods, classes, or programs, there might be different annotation candidates per API call inside the FSpec. In such a situation, we choose the annotation with the highest frequency as the annotation of the API.
    \item Next, using Levenshtein distance metric \cite{levenshtein1966binary}, we calculate dissimilarities between the selected annotations for APIs (\equationref{eq:Levenshtein}) and create the first level of clusters.
\begin{equation}
    \label{eq:Levenshtein}
    lev(a, b)= \left\{\begin{matrix}
    \left | a \right | & \left | b \right | \shorteq \, \shorteq~0\\ 
    \left | b \right | & \left | a \right | \shorteq \, \shorteq~0\\ 
    lev(tail(a), tail(b)) & a\left [ 0 \right ] \shorteq \, \shorteq~ b\left [ 0 \right ]\\ 
    1 + min\left\{\begin{matrix}
    lev(tail(a), b)\\ 
    lev(a, tail(b))\\ 
    lev(tail(a), tail(b))
    \end{matrix}\right. & Otherwise
    \end{matrix}\right.
\end{equation}
    
    In particular, for each branch (i.e., API usage) in FSpec, we first create a list of pairs of APIs, compute the Levenshtein distance score for each pair, and sort the list ascending based on the computed scores. Then, starting from the first element of the list, we create a cluster per pair. Please note that to avoid adding an API in more than one cluster, whenever we select a pair to create a cluster for its corresponding APIs, we remove all the remaining pairs in the list that contain either of the APIs of the selected pair. \ali{Also the Jaro Winkler Distance metric (in code annotator).}   
    \item Finally, we iteratively create hierarchical clusters for APIs in each branch of FSpec based on the distances of clusters' annotations. While forming a cluster, the annotation with the minimum distance to all other annotations in that cluster will be selected as the annotation for the newly created cluster. Due to its hierarchical nature, one would reasonably expect to see the top-most level cluster be annotated as \textbf{\#Authentication}.  \ali{Would they? if the annotation is selected only from the list of API annotations, then, shouldn't the top-level annotation be the same as one of the API annotations?}
\end{enumerate}

\subsection{Step \#2 - Code Annotation}
 \label{sec:IPPS_CodeAnnotation}
Next, we leverage the under-development program as a context, to identify candidate clusters of API calls inside the FSpec that need to be incorporated into the synthesis task. 
To that extent, we find mappings between API clusters inside the FSpec and candidate locations in the program such that a code snippet synthesized from the cluster can be added to the location. Each mapping basically represents \textit{\textbf{what}} part of the tactic should be synthesized, and \textit{\textbf{where}} in the program it should be placed. 
We consider four criteria for creating these mappings, namely, \textbf{(i)} method name similarity, \textbf{(ii)} required variable availability, \textbf{(iii)} correct data- and control-dependencies, and \textbf{(iv)} the quality of the final code. The details of each criterion will be provided later in this section. For each criterion, we compute a score to quantify the likelihood of adding a cluster to that location in the program. We use this score to create a ranked list of candidate pairs of $\langle annotation, location \rangle$ while synthesizing the tactic. 

\subsubsection{Criteria 1: Method Name Similarity}
\label{sec:IPPS_CodeAnnotation_MNS}
The given program might have many methods to which the to-be-synthesized code snippets can be added. Each to-be-synthesized code snippet will be created from API calls inside a cluster. During the FSpec annotation phase (\secref{sec:IPPS_FSpecAnnotation}), we generate annotations for clusters such that each annotation describes the (sub)task that   API calls inside the cluster contribute to. Hence, one heuristic for identifying the proper method of the program for adding a to-be-synthesized code snippet would be to consider methods with a similar name to the cluster's annotation. For the purpose of method name similarity computation, we use \equationref{eq:MethodNameSimilarity} which basically is the inverse of the Levenshtein distance metric that we used before (\equationref{eq:Levenshtein}). Please note that in order to avoid division by zero in case of complete similarity (i.e., $distance=0$), we added a constant (i.e., 1) to the denominator of \equationref{eq:MethodNameSimilarity}. %Moreover, it helps us to better distinguish between names when they are very similar versus the situation when names are less similar. \ali{This needs more explanation.} 
A \textit{method name similarity score (MNS)} is a double number in the range of $(0, 1]$. The higher the MNS score is, the more desirable the method is for adding the to-be-synthesized code snippet to it. \ali{Jaro Winkler Distance metric?} 
\begin{equation}
    \label{eq:MethodNameSimilarity}
    MNS(a, b)= \frac{1}{Lev(a, b) + 1}
\end{equation}
For instance, in the created FSpec, \code{CallbackHandler} instantiation and \code{LoginContext} instantiation are annotated as \textit{\textbf{\#Initialization}}. As shown in the under-development program in \figref{fig:MotivatingExample_UDP}, there is a method named \code{initializeLC(...)}. The name of this method has the most similarity to \#Initialization annotation. The reason is that the MNS score between the ``Initialization'' annotation and the method names ``initializeLC'', ``login'', ``inspectSubject'', and ``main'' is 0.14,  0.09, 0.07, and 0.09, respectively. Therefore, \code{initializeLC(...)} would be at the top of the list of candidate methods for adding the synthesized code snippet from API calls of the \textit{\#Initialization} cluster.

\subsubsection{Criteria 2: Available Variables}
\label{sec:IPPS_CodeAnnotation_VAS}
Each FSpec cluster consists of multiple API calls and their inter-dependencies. If an API call $A_1$ produces data that is used by another API call $A_2$ (i.e., $A_1 \xrightarrow{data} A_2$), it means that $A_2$ is data-dependent to $A_1$. This dependency could be either as a target object $o_1$ (e.g., $o_1 = A_1(...);\ o_1.A_2(...);$), or as an input argument $arg_1$ (e.g., $arg_1=A_1(...);\ A_2(arg_1, arg_2, ...)$). However, there is a possibility that $A_2$ takes another input argument $arg_2$ of type $t_1$ that would not be generated by $A_1$ or any other API calls inside the cluster. Therefore, at least one pre-declared, visible, and initialized variable from type $t_1$ should be \textit{available} in the scope to which the synthesized code snippet from that cluster will be added. This will add another constraint for finding a proper location in the program to synthesize and add each part of the tactic. For instance, annotation \textit{\#Initialization} consists of \code{CallbachHandler} and \code{LoginContext} instantiation. While CallbackHandler does not need any specific input arguments, LoginContext requires an object of type \code{CallbackHandler} and an object of type \code{String}. The instantiated \code{CallbackHandler} would satisfy the need for an argument of type \code{CallbackHandler}. However, still, a variable of type \code{String} should be introduced to the constructor of \code{LoginContext}. Based on this constraint, in the case of the tactic synthesis task demonstrated in \figref{fig:MotivatingExample_UDP}, there are two possible locations in the program to add API calls related to \#Initilization: \textbf{(1)} inside the method \code{initializeLC(...)}, and \textbf{(2)} inside the method \code{main(...)} after line 17. Please note that although there is a variable named \code{varStr} of type \code{String} which is declared at line 2, our approach does not consider this variable since it is not initialized anywhere in the program.

We use JavaParser \cite{danny_van_bruggen_2020_3842713}, which is a popular open-source Java parser tool, for finding visible variables from a queried location. Moreover, to ensure that each of the found variables is initialized before reaching that location, we use the WALA \cite{WALA} program analysis tool to create a context-sensitive data flow for the program. We use \equationref{eq:VariableAvailability} to calculate \textit{variable availability score (VAS)}, which quantifies the suitability of a location $L$ in the program for adding a to-be-synthesized code snippet related to cluster $C$, from a variable availability point of view.

\begin{equation}
    \label{eq:VariableAvailability}
    VAS(C, L) = \left\{\begin{matrix}
    1 & M = 0\\ 
    \prod_{i:1 \to M}^{}min( \frac{Avail(L,T_i)}{Args(C,T_i)}, 1) & M>0
\end{matrix}\right.
\end{equation}

In this equation, $C$ is the cluster, $L$ is the location in the code, and $M$ is the number of different types $T$ that API calls in the cluster $C$ need to receive as unfilled arguments or target objects. If there are no unfilled arguments or target objects in the cluster (i.e., $M=0$), then, the $VAS$ score will be $1$ since the to-be-synthesized code from the cluster $C$ can be added to the location $L$ without a problem (w.r.t., variable availability). Otherwise (i.e., $M>0$) we compute the $VAS$ as explained. For example, in the case of the cluster \textit{\#Initialization}, since the only unfilled argument is of type \code{String}, then, $M$ would be $1$. However, if the argument of type \code{CallbackHandler} was not filled by another API call, $M$ should be considered $2$. 
Moreover, $T_i$ represents the $i$-th type in the cluster, $Avail(L,T_i)$ counts the number of available variables of type $T_i$ at location $L$ in the program, and $Args(C,T_i)$ counts the number of unfilled arguments or target objects in the to-be-synthesized code from cluster $C$ that are of type $T_i$. In case there is at least one distinct available variable of type $T_i$ per needed argument of type $T_i$, the score related to that type will be $1$. Otherwise, it will be a fraction number in the range of $[0, 1)$. If there is no available variable of type $T_i$, then it means that the location $L$ is not suitable for adding the to-be-synthesized snippet from the cluster $C$. The reason is that we can not fill the unfilled arguments of type $T_i$ with a proper variable in this location of the program. The multiplication in the equation guarantees that in such cases the overall score will be $0$. In general, the higher the MNS score is, the more appropriate the location is for adding the to-be-synthesized code snippet. 

%create a 1-CFA \cite{shivers1991control} context-sensitive call graph for the program. A context-sensitive call graph is a directed graph in which the nodes are methods in the program (and third-party libraries used in the program), and edges represent \textit{Caller-Callee} relation between the nodes. In a trade-off between context sensitivity and scalability, we choose 1-CFA and avoid higher sensitivity (i.e., n-CFA for $n>1$). 
%\ali{Needs more technical aspects related to JavaParser and Dataflow-analysis?}.

\subsubsection{Criteria 3: Control- and Data-dependencies}
\label{sec:IPPS_CodeAnnotation_CDS}
From the previous two criteria, we were able to identify and rank candidate locations in the program for synthesizing code snippets related to a cluster. Those two criteria focus on API calls inside each cluster. However, in addition to control- and data-flow dependencies between API calls inside a cluster (i.e., intra-cluster dependencies), there are some dependencies between API calls of different clusters (i.e., inter-cluster dependencies) that need to be considered while searching for proper locations in a program.
In other words, the synthesizer should also preserve the correct control- and data-flow dependencies between API elements of different clusters during the synthesis process. 
%This criteria is responsible for correct orchestration of synthesized code snippets in the program.
For instance, let's assume that the synthesizer adds the API \code{login()} of class \code{LoginContext} which belongs to cluster $C_2$ to a location \textit{$L_2$} in the program. In such a case, the synthesizer has to add the \textit{initialization} of \code{LoginContext} which belongs to cluster $C_1$ to a location \textit{$L_1$} such that \textit{$L_1$} is visited before \textit{$L_2$} while running the program. 

In order to check the above condition, we first perform an inter-procedural context-sensitive static analysis over the program to find data- and control- dependencies between different \textit{scopes} of the program. We consider the basic blocks of the control-flow analysis of the program as the scope in our approach. Next, based on the performed analysis, we create a scope dependency graph in which the nodes are identified scopes, and the edges are data- and control dependencies between scopes. Please note that each to-be-synthesized code snippet (created from API calls in a cluster) will be placed in a scope. In fact, by selecting a candidate location for a cluster, we basically annotate a node in the scope dependency graph with the annotation of that cluster. If we keep the annotated nodes and remove the remained nodes from the graph (and yet preserve the direct and indirect edges between the remained nodes), the final graph would represent a program-wide control- and data-dependencies between to-be-synthesized code snippets. In order to have a correct tactic synthesis, this graph should look the same as the sub-graph of the FSpec where the selected clusters belong to. If this constraint is satisfied, then, it means that the selected mapping between clusters and locations will result in a correct tactic implementation. Otherwise, it means that the combination of selected locations for clusters was not correct and thus, the selected mapping will be disregarded. In such a case, the next best candidate mapping from the ranked list of mappings will be considered for the next round of control- and data-dependency checking. The result of this analysis will be the \textit{Control- and Data-dependency Score (CDS)} for the entire mapping, which is a binary number $0$ or $1$. If correct dependencies are observed, then the score will be $1$. Otherwise, it will be $0$.

\subsubsection{Criteria 4: Synthesized Code Quality}
\label{sec:IPPS_CodeAnnotation_CQS}
Lastly, we would like to have the tactic implemented in a way that the final code has a high quality. There are a variety of techniques that quantify the quality of the code from different perspectives, e.g., high-cohesion and loose-coupling as two fundamental quality metrics for code \cite{yourdon1979structured}. In this paper, we focus on \textit{high-cohesion}, but we will discuss how \textit{loose-coupling} could also be incorporated as future work. 

High-cohesion and loose coupling can be measured based on considering different scope granularities, including method-level, class-level, package-level, and module-level. In this paper, without affecting the generalizability of the approach, we focus on class-level granularity, meaning that we intend to have classes where single functionalities are implemented in separate methods to improve the \textit{cohesion}, while functional methods (i.e., methods that implement functionalities) have low dependencies to each other to reduce the \textit{coupling}. However, one can easily change the granularity of the scope, e.g., consider package-level or module-level scopes. 

The general idea is to give a higher score to the implementation that adds functionality-relevant code (i.e., lines of code that collaboratively implement a single functionality) to the same method and places those that implement different functionalities in different methods (i.e., high-cohesion). Also, to keep a loose coupling, one would keep the (direct) interaction between these methods as minimum as possible.

As we discussed before in \secref{sec:IPPS_FSpecAnnotation}, \textsc{IPSynth} considers API calls that collaborate to implement a single functionality in a cluster. Also, \textsc{IPSynth} always generates the corresponding code to a cluster of API calls as a block of code, keeping the code inside this block always together, and putting them in the same method. Therefore, it always generates and adds the lines of code that implement a single functionality in the same method. However, to guide the synthesizer to put the synthesized code for different clusters in separate methods (to avoid violation of the single responsibility rule and improve the code cohesion), we use Code Quality Score (CQS) shown in \equationref{eq:code_quality_score} that measures the cohesion level of the code. 

\begin{equation}
    \label{eq:code_quality_score}
 CQS = \frac{1}{n} \sum_{i=1}^{n} \frac{1}{Clst(m_i)} 
\end{equation}

In this equation, $n$ is the number of the methods that the approach puts the synthesized code snippets in, $m_i$ is a method that at least one synthesized code snippet is placed in, and the $Clst$ function counts the number of code snippets that were placed in method $m_i$. Aiming for a high $CQS$ helps \textsc{IPSynth} to synthesize high-quality tactic implementation (w.r.t. high-cohesion). The highest possible score is $1$ where each code snippet is placed in a separate method. On the other side, the more functionality-irrelevant code snippets placed in a single method, the closer this score can get to $0$. 

For example, let's assume that there are two methods $m_1$ and $m_2$ where there are two code snippets synthesized and placed in $m_1$ and one code snippet synthesized and placed in $m_2$. Based on the introduced $CQS$, the quality code score for such a synthesis task will be $\frac{1}{2}(\frac{1}{2} + \frac{1}{1}) = 0.75$. Please note that this score only takes care of the code quality. For example, if there is a method in the program that there should be a synthesized code snippet placed in it but it is not, then, the other metrics (Criteria 1 and 2) will take care of this issue.

The introduced quality metric (\equationref{eq:code_quality_score}) focuses on high-cohesion, however, it would be possible to extend it with \equationref{eq:code_coupling_score} that incorporates the coupling status of the code and computes the code \textit{instability} \cite{martin2003agile}.

\begin{equation}
    \label{eq:code_coupling_score}
 CCS = \frac{1}{n} \sum_{i=1}^{n} \frac{Ce_{m_i}}{Ce_{m_i} + Ca_{m_i}} 
\end{equation}

In this equation, $n$ is the number of the methods that the approach puts the synthesized code snippets in, $m_i$ is a method that at least one synthesized code snippet is placed in, $Ce_{m_i}$ is the \textit{efferent} coupling, i.e., the number of other methods that method $m_i$ depends upon, and $Ca_{m_i}$ denotes the \textit{afferent} coupling, i.e., the number of other methods that depend on method $m_i$. A $CCS$ score close to $0$ indicates a stable method with fewer dependencies, while a value close to $1$ shows a high dependency on the other methods. 

\subsubsection{Putting All Together}
\label{sec:IPPS_CodeAnnotation_altogether}
We incorporate all four constraints above to spot the proper locations in the code for creating and adding to-be-synthesized codes to the program. More specifically, we use the first two scores (MNS and VAS) to generate a cluster-location score (CLS) to find and rank appropriate locations per cluster. \equationref{eq:Cluster_Location_Score} demonstrates how we generate this score for each mapping. 

\begin{equation}
    \label{eq:Cluster_Location_Score}
CLS=
\frac{(c_{MNS} \times MNS) + (c_{VAS} \times  VAS) }{c_{MNS}+c_{VAS}}
\end{equation}

In this equation, $MNS$ and $VAS$ are methods naming similarity and variable availability scores computed before. In addition, $c_{MNS}$, and $c_{VAS}$ are coefficients for $MNS$ and $VAS$. These coefficients denote the effect of each quantified criterion in the final score. For the sake of simplicity, we consider all the coefficients as $1$ in our computations. However, in future work, it would be possible to leverage a training procedure using a training dataset to statistically learn each coefficient. The CLS score quantifies the worthiness of a location for adding a synthesized cluster to that location in the program. However, a tactic synthesis process is composed of a couple of cluster synthesis tasks. For a successful and correct tactic synthesis, we need to make sure that (i) the final code has high quality and (ii) there are correct control- and data dependencies between synthesized cluster-related code snippets in the final code. Considering all the selected mappings, \equationref{eq:CodeAnnotationScore} generates a score for the entire mappings in the synthesis task. 

\begin{equation}
    \label{eq:CodeAnnotationScore}
CAS = %CDS \ ( \frac{1}{n} \sum_{i=1}^{n} CLS_i )
\left\{\begin{matrix}
0 & CDS=0\\ 
\frac{(c_{CQS} \times CQS + c_{CLS} \times ( \frac{1}{n} \sum_{i=1}^{n} CLS_i ))}{c_{CQS} + c_{CLS}}  & CDS=1
\end{matrix}\right.
\end{equation}

%     VAS(C, L) = \left\{\begin{matrix}
%     0 & M = 0\\ 
%     \prod_{i:1 \to M}^{}min( \frac{Avail(L,T_i)}{Args(C,T_i)}, 1) & M>0
% \end{matrix}\right.

In this equation, $CLS_i$ is the computed $CLS$ score for the $i^{th}$ mapping in the synthesis task, $n$ denotes the total number of mappings in the synthesis task, $CQS$ is the code quality score for all the clusters incorporated in a synthesis task, and $CDS$ is the control- and data-dependency score computed for the mappings in the synthesis task. Moreover, $c_{CQS}$ and $c_{CLS}$ are coefficients for applying the effect of each of the $CQS$ and $CLS$ in the final score. Based on our observation, it would be very helpful if the $c_{CQS}$ is very low, just enough to make a difference when the $CLS$ scores of two mappings are equal. Throughout our experiments, we found $0.0001$ to be an efficient value for $c_{CQS}$ while $c_{CLS}$ is set to $1$.  
Furthermore, in case the control- and data-dependency criteria is not satisfied (i.e. $CDS = 0$), the formula completely disqualifies the list of selected mappings and returns $0$. In such a case, the synthesizer goes over the ranked list of the locations for clusters and tries to find substitute mappings such that the combination of $CLS$ and $CQS$ is the maximum possible value while the $CDS$ is not $0$.

%satisfied (i.e., $CDS = 1$), this formula computes the worthiness of the mappings as the average of CLS scores of all the selected mappings multiplied by its quality score (between $0$ and $1$). Otherwise 

Please note that considering only correct locations for synthesizing clusters will result in a \textbf{correct by construction} approach. In other words, by following this algorithm, we make sure that the final synthesized tactic is implemented correctly.
%\equationref{eq:CodeAnnotationScore} helps us to only accept mappings that preserve this constraint as well.
%we compute a two-level code annotation score (CAS) which quantifies the worthiness of a list of mappings between clusters and locations. In the first level, we assign a score to a given mapping $<cluster, location>$. 

%the code annotation score (CAS) for a given mapping between a location in the program for synthesizing and adding a specified annotation from \equationref{eq:CodeAnnotationScore}. This would be a score ranging from 0 to 1.
%

%
%In this equation, $MNS$, $VAS$, and $CDS$ are scores computed in \secref{sec:IPPS_CodeAnnotation_MNS}, \secref{sec:IPPS_CodeAnnotation_VAS}, and \secref{sec:IPPS_CodeAnnotation_CDS}, respectively. Also, $c_{MNS}$, and $c_{VAS}$ are coefficients for $MNS$ and $VAS$. These coefficients denote the effect of each quantified criteria in the final score. Please note that if $CDS$ is 0, then, it means that the expected control- and data-dependencies inside annotations have not been preserved. Therefore, no matter what $MNS$ and $VAS$ scores are, this location should not be considered for the given FSpec annotation at all. Hence, the value of $CAS$ in such a case would be 0. Otherwise, it would be a function of $MNS$ and $VAS$.
%For the sake of simplicity, we consider all the coefficients as $1$ in our computations. However, as a future work, it would be possible to leverage a training procedure using a training dataset to learn each coefficient. 

\ali{Create an example of how the entire approach works, from annotating the FSpec, to annotating the code through the given formula, to generating sketches, and finally, resolving the sketches.}

\subsection{Step \#3 - Sketch Generation}
\label{sec:IPPS_SketchGeneration}
\begin{figure}[t]
    \centering
    \includegraphics[width=.9\linewidth]{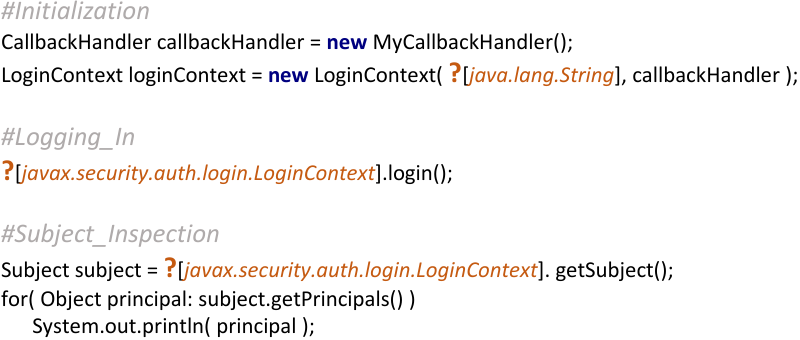}
        \centering
    \caption{Sketches generated for \#Initialization, \#Logging\_In, and \#Subject\_Inspection annotations.}
    \label{fig:sketchGeneration}
\end{figure}

Through the previous step, we identified candidate locations in the program per cluster such that the to-be-synthesized code snippet for that cluster will be placed in that location. Each cluster consists of a set of APIs and their dependencies (i.e., control and data dependencies) represented as a graph in which the nodes are API calls and the edges are their dependencies. In this step, we translate this graph into a code snippet that is only composed of those API calls and generate the structure of the to-be-synthesized code snippet. This code snippet is in the format of a Static Single Assignment (SSA). For instance, if there is a data dependency between API \code{$A_1$} and \code{$A_2$} in a cluster such that $A_2$ uses the data generated by $A_1$ (i.e., \code{$A_1$} $\xrightarrow[]{data}$\code{$A_2$}), the output of \code{$A_1$} would be stored in a variable $v_1$ such that $v_1$ is an input argument to or a target object for \code{$A_2$} in the corresponding generated code snippet. However, there are still two  dependencies remained to be added to this code snippet, \textbf{(i)} dependencies between the code snippet and variables available (visible and initialized) at the location in which the code snippet will be added to, and \textbf{(ii)} dependencies between this code snippet and the other generated code snippets which reflects the dependencies between annotated clusters. To address these dependencies, in this step, we put \textbf{\textit{holes}} in the generated code snippet and we will fill in these holes later in the next step (\secref{sec:IPPS_SketchResolving}). Each hole is a placeholder for a variable in the code snippet. In other words, while translating a graph to a code snippet, if an argument or a target object is not filled with an API from the same graph,
we put a hole in that part of the code snippet. The output of this step would be a \textbf{\textit{sketch}} of the final version of  the to-be-synthesized code snippet for each cluster. For instance, \textit{\#Initialization}, \textit{\#Logging\_In}, and \textit{\#Subject\_Inspection} annotations will be translated into the sketches shown in \figref{fig:sketchGeneration}.
In this figure, question marks (i.e., \textbf{?}) are the holes needed to be filled out by variables available at the location where the sketch will be added to. Note that from the FSpec, we know the signature of all APIs inside a cluster. Therefore, wherever we put a hole in the constructed sketch, we are also aware of the type of variable that should be used to fill in that hole. For instance, the FSpec provides the signature of the constructor of \code{LoginContext}, including the types of its input arguments. Hence, we can infer that the type of hole in the constructor of \code{LoginContext} in \figref{fig:sketchGeneration} is \code{java.lang.String}. Also, holes available in sketches created for \textit{\#Logging\_In} and \textit{\#Subject\_Inspection} are of type \code{javax.security.auth.login.LoginContext}.

\subsection{Step \#4 - Sketch Resolution}
\label{sec:IPPS_SketchResolving}
So far, we have created sketches of to-be-synthesized code snippets. These sketches contain holes that need to be filled in by available variables. From one side, we know the type of each hole, and from another side, there are a limited number of variables available with the same type as the hole's type. This is a constraint-based problem in which we need to find a solution that satisfies all the constraints. In this step, based on the mentioned constraints, we translate the problem of finding proper variables for filling in the holes in sketches to a Satisfiability Modulo Theory (SMT) problem. Then, we use an off-the-shelf SMT solver, Z3 \cite{de2008z3}, to find a proper solution that satisfies all the constraints. Finally, we translate back the found SMT solution to a concrete solution for the original problem of finding proper variables for holes. 

To do this, we first collect information about all the available and initialized variables that were identified through Step 2 to create a ranked list of possible variables for each hole. The closer the variable is to the location of a hole in the code, the higher its position would be in the ranked list. Let us assume that for each hole $h_i$, there would be a list of $n$ possible variable names $[v_{i,1}, v_{i,2}, ..., v_{i,n}]$ to choose from. The final goal is to find a minimum number of variables such that all the holes could be filled with \ali{Why minimum number is important?}. We first translate the sketch problem to the following SMT problem:

\begin{equation}
    \label{eq:SketchResolving1}
    \begin{split}
Constraint_1 = (v_{1, 1}\vee v_{1, 2} \vee v_{1, 3} \vee ... \vee v_{1, m_1}) \wedge \\(v_{2, 1}  \vee v_{2, 2} \vee  v_{2, 3} \vee ... \vee v_{2, m_2})  \wedge \\... \wedge (v_{n, 1}\vee v_{n, 2} \vee v_{n, 3}  \vee ... \vee v_{n, m_n})
\end{split}
\end{equation}

where $v_{i,j}$ is a Boolean variable that, when true, indicates that the $j$-th element in the list of available variables for hole $h_i$ has been selected. Please note that holes' variable lists are not mutually exclusive, meaning that it is possible that a variable is in the ranked list of more than one hole. Let us say that the $i$-th variable of the ranked list of hole $h_1$ and the $j$-th variable of the ranked list of hole $h_2$ are the same. Moreover, the $k$-th variable of the ranked list of hole $h_1$ and the $s$-th variable of the ranked list of hole $h_3$ are the same. In such a case, we have another constraint as shown in \equationref{eq:SketchResolving2}.  

\begin{equation}
    \label{eq:SketchResolving2}
    \begin{split}
Constraint_2 = (v_{1, i} \iff v_{2, j}) \wedge (v_{1, k} \iff v_{3, s})
\end{split}
\end{equation}

where the iff operator (i.e., $\iff$) implies that the value of its operands should be the same. Since we want to satisfy all the constraints at the same time, the final SMT problem would be the conjunction of $Constraint_1$ and $Constraint_2$, i.e., $Constraint_1 \wedge Constraint_2$.
Please recall that the goal of this step is to use the minimum number of available variables for filling out all the holes in the sketch. Therefore, we ask the SMT solver to find the minimum number of boolean variables such that $Constraint_1 \wedge Constraint_2$ can be $True$. 
\section{Evaluation}
\label{sec:IPSynth-experimentalStudy}
This section details our experimental study of \textsc{IPSynth} for synthesizing tactic codes and adding them to a program. Moreover, it compares the synthesis results of \textsc{IPSynth} against a powerful code generator tool, ChatGPT \cite{OpenAI2023GPT4TR}. 

For the first time, we have created a dataset of architectural tactic synthesis tasks to be used to evaluate program synthesis techniques. This dataset was used as part of our experimental study in this paper. To evaluate the introduced approach, we conduct two types of evaluations: \textbf{(i)} a fine-grained evaluation of the components (i.e. steps) of our approach (\secref{sec:eval_component}), and \textbf{(ii)} an evaluation of the performance of the entire approach (\secref{sec:eval_overall}). While in the former we investigate the performance and correctness of each component, in the latter, we inspect the output of the entire approach. We also compare the generated code by \textsc{IPSynth} against one of the most powerful related works that is an expert in code generation, ChatGPT \cite{OpenAI2023GPT4TR}. \ali{and maybe Google Bard?} In the rest of this section, we first introduce the dataset, then, we discuss the experiments and their results, and finally, we compare the results of our approach against the related work. 

\subsection{Architectural Tactic Synthesis Dataset}
\label{sec:eval_dataset}
\begin{figure}
    \centering
    \includegraphics[width=.9\textwidth]{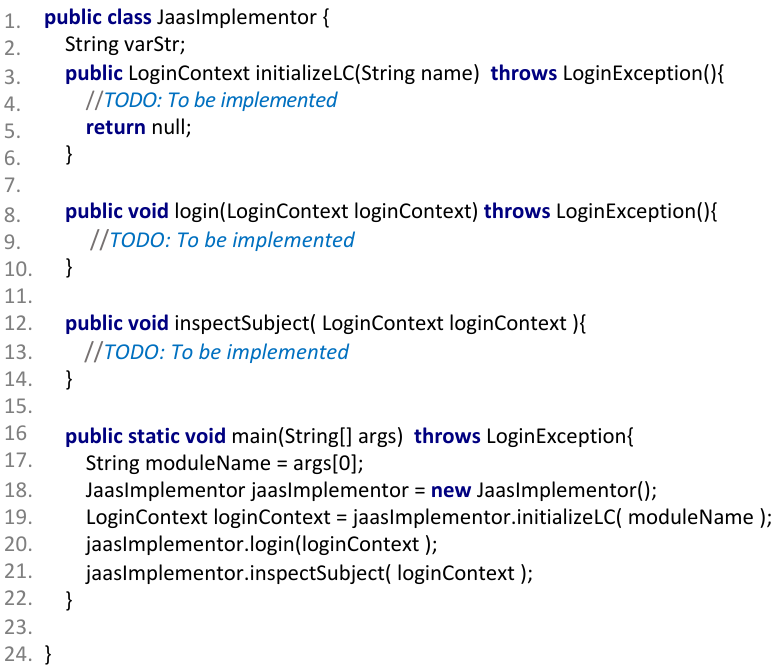}
    \caption{A sample test program for the tactic synthesis task.}
    \label{fig:IPSynth-Dataset-TestProgram}
\end{figure}

\begin{figure}
    \centering
    \includegraphics[width=.85\linewidth]{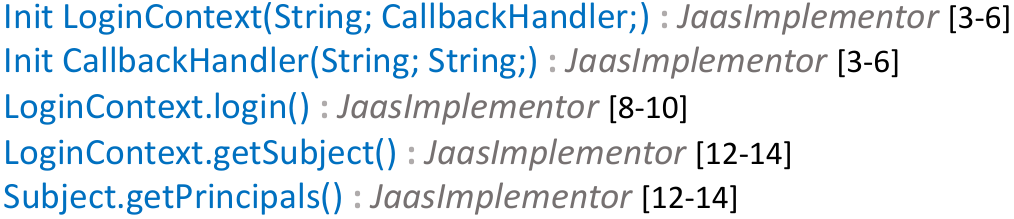}
        \centering
    \caption{A test label for the test case shown in \figref{fig:IPSynth-Dataset-TestProgram}.}
    \label{fig:eval_dataset_label}
\end{figure}

We created a dataset of architectural tactic synthesis tasks which includes test programs and their related labels. Each task represents a unique way of implementing a \textit{security} tactic. The synthesizer is allowed to use JAAS API for synthesizing the tactic. The current version of the dataset consists of \textbf{20} programs. Each curated test program is syntactically correct, and there are places in the program where the synthesizer is expected to generate tactic pieces and add them to the mentioned locations. The correctness of test programs was confirmed by the Java compiler (syntax correctness) as well as the review team where the members were expert programmers and completely familiar with the JAAS framework. \figref{fig:IPSynth-Dataset-TestProgram} shows such a program where tactic codes are expected to be implemented in three separate methods that inter-procedurally collaborate to execute the tactic. Since we know where tactic pieces should be added, we were able to create a data structure that represents the expected locations of tactic pieces; we used that information as the label for each of the created test cases. \figref{fig:eval_dataset_label} presents the created label for the test program provided in \figref{fig:IPSynth-Dataset-TestProgram}. Each label is in the format of a dictionary where the \textit{key} is the tactic piece and the value is the possible lines in a file that the synthesized code is expected to be added such that the implementation is correct. In our experiments, we use this information to assess the correctness of the synthesized code. 


All the test cases are inter-procedural tasks, meaning that there are more than one method (in average $4.5$ methods) in a test case that collaborate to accomplish the tactic implementation task. We made this dataset publicly available at \url{https://anonymous.4open.science/r/Anonymous-82DE}. \tableref{tbl:IPSynth_eval_comparison_dataset} provides a summary of this dataset. In this table, programs are numbered from $1$ to $20$. For each program, we provide its corresponding information including the number of methods in the program, number of global variables that might affect the synthesis process, number of classes, number of java files, and finally, a description of the complexity of each program compared to its previous program. Please note that the complexity of programs increases by their numbers, i.e., implementing a tactic in $P10$ is more complicated compared to $P5$. 

In addition to the evaluation of the performance of \textsc{IPSynth} on the introduced dataset, we also studied the performance of \textsc{IPSynth} in comparison with one of the most powerful tools in code synthesis, ChatGPT\ali{and Google Bard?}. This study gives us a better understanding of the advantages of our semi-formal approach against LLM-based approaches such as ChatGPT. Moreover, we learned valuable lessons about the possibility of mixing the strength of \textsc{IPSynth} and LLM-based approaches for improving the results. We discuss these findings in this section (\secref{sec:IPSynth_eval_comparative_analysis}), as well as in the future directions of the work (\secref{sec:Conclusion}). 

\begin{table*}
\centering
\caption{Description of the tactic synthesis dataset}
 \label{tbl:IPSynth_eval_comparison_dataset}
\begin{tabular}{P{0.05\linewidth} P{0.10\linewidth} P{0.06\linewidth} P{0.07\linewidth} P{0.06\linewidth} p{0.5\linewidth}} 
    \toprule
 Prog. No. & \# of methods & \# of global vars & \# of classes & \# of files & Added complexity \\ 
    \midrule

    \textbf{P1} & 4 & 1 & 1 & 1 & \\
    \textbf{P2} & 4 & 1 & 1 & 1 & Nested blocks \\
    \textbf{P3} & 4 & 4 & 2 & 1 & An inner class \\
    \textbf{P4} & 5 & 4 & 2 & 1 & More methods \\
    \textbf{P5} & 5 & 2 & 1 & 1 & Two non-initialized global vars in the main class \\
    \textbf{P6} & 5 & 2 & 1 & 1 & A global and a static non-initialized string var \\
    \textbf{P7} & 6 & 2 & 1 & 1 & More methods \\
    \textbf{P8} & 6 & 2 & 1 & 1 & Different types of argument passing \\
    \textbf{P9} & 7 & 1 & 1 & 1 & A global initialized double var \\
    \textbf{P10} & 8 & 1 & 1 & 1 & Nested method calls \\
    \textbf{P11} & 1 & 1 & 1 & 1 & Non-initialized global field \\
    \textbf{P12} & 2 & 1 & 1 & 1 & A not-called method \\
    \textbf{P13} & 6 & 1 & 2 & 1 & A non-initialized boolean field + an inner class \\
    \textbf{P14} & 7 & 4 & 2 & 1 & Nested method calls + an inner class with a private method. \\
    \textbf{P15} & 8 & 7 & 3 & 1 & Nested method calls + 2 inner classes, both with private methods \\
    \textbf{P16} & 9 & 7 & 4 & 2 & Two java files in the same package, 2 inner classes with private methods, nested method calls \\
    \textbf{P17} & 11 & 7 & 5 & 3 & Three java files in the same package, 2 inner classes with private methods, nested method calls \\
    \textbf{P18} & 14 & 8 & 6 & 4 & Four java files in 2 different packages, 2 inner classes with private methods, nested method calls \\
    \textbf{P19} & 17 & 10 & 8 & 5 & Five java files in 3 different packages, 3 inner classes with private methods, bested method calls \\
    \textbf{P20} & 21 & 15 & 10 & 6 & Six java files in 4 different packages, 4 inner classes with private methods, nested method calls  \\

    \bottomrule
\end{tabular}
\end{table*}

\subsection{Evaluating Approach Components}
\label{sec:eval_component}
We first investigate each component of the approach to find a better insight into the  performance of that component. This would help us pinpoint possible bottlenecks as well as the strength points of the approach.
\subsubsection{Step 1: FSpec Annotation}
\ali{This needs another round of running experiments. }There are two main criteria for the evaluation of FSpec annotation (introduced in \secref{sec:IPPS_FSpecAnnotation}), \textbf{(i)} clustering API elements in the FSpec, and \textbf{(ii)} selecting a proper label for each cluster. For this purpose, the review team reviewed all clusters and their associated annotations. The team members are familiar with security tactics, especially with the JAAS framework.

\textit{- API Clustering:}
The built framework API specification model (FSpec) has 88 nodes, i.e., API elements, and 130 edges, i.e., dependencies. Following a similar approach discussed in \secref{sec:IPPS_FSpecAnnotation}, we created clusters of APIs. Each cluster contains a set of API calls that contribute to a functionality. 

\textit{- Label Generation:}
We also followed a similar approach to what was discussed in \secref{sec:IPPS_FSpecAnnotation} and generated labels (i.e., annotations) for clusters. The review team then verified the quality of generated annotations based on two metrics: \textbf{(i)} the name similarity metric between API names in a cluster and its generated annotation, as well as \textbf{(ii)} a manual review process that the review team performed to assess the relevancy of the generated annotations with the name of the methods in which APIs within the cluster can be placed in. 

\subsubsection{Step 2: Code Annotation}
As discussed before in \secref{sec:eval_dataset}, we inspected all the test cases and created a list of correct locations for each of the tactic pieces to be synthesized and added to. 
%For instance, the label shown in \figref{fig:eval_dataset_label} indicates that the \textit{\#Initialization} annotation could be synthesized and added to each of lines 3-6 and 17-22. Moreover, \textit{\#Login} and \textit{\#SubjectInspection} annotations could be synthesized and added to lines [8-10, 19-22] and [12-14,19-22], respectively. Please note that since these pieces have control- and data-dependencies, selecting a location for each piece affects the choices for the other pieces. 
%
In the following experimental studies, we measure the accuracy of the returned recommendations using two metrics that are widely used in the recommendation tasks, \textit{Hit Ratio@K} (HR@K) and \textit{Mean Reciprocal Rank (MRR)}. Please note that due to the different number of possible correct answers per test case, measuring $Recall@K$ and $Precision@K$ would not be an option in our experimental studies. Therefore, we do not include these metrics in our experimental studies. \ali{Does it make sense? }To calculate the HR@K, we use the following equation:

\begin{comment}
\begin{equation}
    \label{eq:top_k_accuracy}
Top\mbox{-}k\ Precision=\frac{100}{n*k}\sum_{i:1 \rightarrow n}^{}\sum_{j:1 \rightarrow k}^{}[isCorrect(M_{i,j})\ ?\ 1\ :\ 0] 
\end{equation}
\end{comment}

\begin{equation}
    \label{eq:hit_ratio_k}
HR@K=\frac{100}{n}\sum_{i=1}^{n}(hasCorrect(M_{i}[1:k])\ ?\ 1\ :\ 0) 
\end{equation}

where $k$ denotes the cutoff point of the ranked list of suggested mappings returned by our approach, $n$ represents the number of test cases in the evaluation process, $M_{i}[1..k]$ is the sub-list of the ranked recommendations for test case $i$ from element $1$ to the  $k$th element, and finally, $hasCorrect(...)$ checks whether any of the mappings in the given list are amongst the correct mappings. For instance, if $k=1$ (i.e., top-1), we only consider the first mapping in the ranked suggestion list returned by our code annotator and investigate whether that mapping is correct. Likewise, if $k=5$, it means that we let the code annotator return a ranked list of $5$ suggestions and we search for a correct mapping among the returned list. Moreover, we use \equationref{eq:mrr} to calculate the MRR metric for the list of recommendations.

\begin{equation}
    \label{eq:mrr}
MRR=\frac{1}{n}\sum_{i=1}^{n}(\frac{1}{rank_i}) 
\end{equation}

In this equation, $n$ denotes the number of test cases, and $rank_i$ refers to the rank position of the first correct mapping in the returned mapping list. Please note that, unlike HR@K, in which we find accuracy based on different cutoffs (i.e., $k$), we consider the entire recommended list while computing the MRR of the recommendations. In our experiments, we consider the maximum length of the ranked list as $100$. Therefore, if a correct answer was not found in the ranked list, the $\frac{1}{rank_i}$ part in \equationref{eq:mrr} would be considered as $0$.

\begin{table*}
\centering
\caption{Code annotation Hit Ratio for Top-K suggestions (HR@K) per scoring method.}
    \label{tbl:eval_codeannotation}

\begin{tabular}{cccccccccc} 
\toprule 
 \multirow{2}{*}{Criteria} & \multicolumn{8}{c}{HR@K} & \multirow{2}{*}{MRR}\\ 

  & $k=1$ & $k=2$ & $k=3$ & $k=4$ & $k=5$ & $k=10$ & $k=50$ & $k=100$ & \\  
    \midrule
 MNS & 80\% & 80\% & 80\% & 80\% & 80\% & 80\% & 80\% & 90\% & 0.75\\ 
 AVS & 15\% & 20\% & 20\% & 20\% & 20\% & 45\% & 50\% & 85\% & 0.21 \\ 
 CQS & 5\% & 5\% & 5\% & 5\% & 5\% & 5\% & 5\% & 10\% & 0.05 \\ 
 CAS & \textbf{85}\% & \textbf{90}\% & \textbf{90}\% & \textbf{90}\% & \textbf{90}\% & \textbf{90}\% & \textbf{90}\% & \textbf{100}\% & \textbf{0.88}\\ 
 \bottomrule
\end{tabular}
\end{table*}

\textit{- Method Name Similarity:} We first only consider the MNS score (\secref{sec:IPPS_CodeAnnotation_MNS}) to find the correct spots in the program to synthesize each FSpec annotation. We implemented the Levenshtein distance similarity through dynamic programming, which significantly reduces the run-time of the MNS computation process. The first row of \tableref{tbl:eval_codeannotation} shows the result of this experiment. When we only incorporate the MNS score to identify locations, on average, only 80\% of the top-1 locations suggested by the approach point to the correct spots in the program for the synthesis task. When extending the accepted suggestions to top-100, 90\% of the returned suggestions are correct. Moreover, the MRR of the returned suggestions is $0.75$. 
% This basically means that by only using the MNS metric, the correct answer (if available) can be found between the second and third returned elements of the ranked list. 
Based on this experiment, relying only on the method naming similarity score (MNS) would not result in a precise location spotting in our approach.

\textit{- Available Variable:}
Next, we only consider the AVS score (discussed in \secref{sec:IPPS_CodeAnnotation_VAS}) for finding the correct locations in the code for the tactic synthesis task. The second row of \tableref{tbl:eval_codeannotation} presents the result of this experiment, which shows a worse result compared to the situation in which we only use the MNS score. For example, the hit ratio of top-1 dropped from 80\% to 15\%. The main reason is that in each program there are a couple of methods that might provide all the required available variables for a code piece. Therefore, relying only on AVS can lead to confusion for the synthesizer and it might choose an undesired method at the end. Also, the MMR for this experiment is $0.21$, meaning that if available, the correct suggestion can be found between the fifth and sixth mappings in the recommended list.

\textit{- Code Quality:}
In the next experiment, we consider the effect of code quality alone, without other metrics. In this case, we assume that the MNS score, as well as the AVS score for all the mappings, are both $1$, meaning that we do not bring their effect into the code annotator. As shown in the third row of the \tableref{tbl:eval_codeannotation}, the $CQS$ cannot provide insightful guidance for the synthesizer, however, as we discuss later, it can contribute to better annotation when used with the other metrics.

\textit{- Control- and Data-dependency:}
We do not provide a separate evaluation for the Control- and Data-dependency Score (CDS). The reason is that this score basically quantifies the worthiness of a mapping that has been created based on MNS, CQS, and AVS scores. Therefore, we do not generate a list of ranked  suggestions only based on this score. However, the effect of CDS would be observable when all the scores are incorporated (all together) as discussed in the next item.     

\textit{- All Together:} 
Finally, we incorporate all the code annotation score (CAS) metrics introduced before for finding the correct locations in the code for the synthesis task. The fourth row of \tableref{tbl:eval_codeannotation} presents the result of this experiment. The result shows that the hit ratio for the top-1 and top-2 suggestions are 85\% and 90\% respectively when considering the CAS score (All scores together). This is a significant improvement compared to using each of the metrics individually. In addition, the MMR for this metric is $0.88$ which means that the correct suggestion if available, would most likely be the first element in the ranked list. The results suggest that the code annotation metric we introduced in this paper (i.e., CAS) enables the synthesizer to identify the code locations for the synthesis task with high accuracy.

\subsubsection{Step 3: Sketch Generation}
Each sketch is a template of a to-be-synthesized code snippet which will be a part of the to-be-implemented architectural tactic. Hence, we investigate the accuracy of sketches generated by our approach based on three criteria: \textbf{(i)} incorporation of a correct set of API calls in the sketch, \textbf{(ii)} the correctness of types associated to each hole in the sketch, and \textbf{(iii)} establishment of proper data- and control-dependencies between APIs inside the sketch. For this purpose, the review team reviewed all the generated sketches. \tableref{tbl:eval_sketchgeneration} provides a summary of this investigation. More details are provided as follows: 

\begin{table}
\centering
\caption{Sketch generation results \ali{double check}}
 \label{tbl:eval_sketchgeneration}
\begin{tabular}{P{0.25\linewidth} P{0.15\linewidth} P{0.2\linewidth} P{0.15\linewidth} } 
    \toprule
 Criteria & \# of cases & \# of correct answers & Accuracy \\ 
    \midrule
 Set of APIs & 50 & 50 & 100\% \\ 
 Hole types. & 60 & 60 & 100\% \\ 
 Control dep. & 50 & 50 & 100\% \\ 
 Data dep. & 60 & 60 & 100\% \\ 
  \bottomrule
\end{tabular}
\end{table}

\textit{- Correct set of APIs:}
%Below line could be removed
There are specific API elements in each FSpec cluster that must be appeared in the corresponding sketch. These API calls are basic building blocks of the to-be-synthesized pieces. 
We inspected all the generated sketches to identify if they contain all the API elements from their corresponding FSpec clusters. There were 50 API calls identified in clusters. We were able to confirm that all API calls (100\%) from FSpec clusters appeared in their corresponding sketches.  

\textit{- Correct types for holes:}
There were $60$ holes generated for the test cases in the experimental study. These holes were of different types (e.g. \code{javax.security.auth.login.LoginContext}, \code{java.lang.String}, etc.). We manually inspected the associated types of each hole. All holes were associated with the correct type. Hence, the accuracy of the type assignment for the holes was 100\%.

\textit{- Correct control- and data-dependencies:}
We also investigated the correctness of established inter-dependencies between APIs of sketched code snippets. Overall, there were 50 control-, and 60 data-dependencies. All the established dependencies were identified as correct dependencies. Therefore, the accuracy in establishing correct control- and data-dependencies inside each sketched code snippet was 100\%.

\subsubsection{Step 4: Sketch Resolving}
The expected output of the sketch resolving step (\secref{sec:IPPS_SketchResolving}) is a program without holes. Moreover, the program is supposed to be syntactically error-free, i.e., the program should be compilable. \tableref{tbl:eval_sketchresolving} shows the result of this experiment. We provide more details as follows:  
\begin{table}
\centering
\caption{Sketch resolving results \ali{Double check this number.}}
 \label{tbl:eval_sketchresolving}
\begin{tabular}{P{0.3\linewidth} P{0.18\linewidth} P{0.2\linewidth} P{0.15\linewidth} } 
    \toprule
 Criteria & \# of cases & \# of correct answers & Accuracy \\ 
    \midrule
 Resolving holes & 60 & 60 & 100\% \\ 
 Compilable & 20 & 20 & 100\% \\ 
  \bottomrule
\end{tabular}
\end{table}

\textit{- Resolving all the holes:}
Our synthesizer was able to resolve all 60 generated holes, which means that exactly one variable was assigned for each hole. In addition, the type of all the assigned variables was matched with the type of their corresponding holes. 

\textit{- No syntax error:}
Moreover, at the end of the synthesis process, we inspected the synthesized programs (20 programs) to assess their syntax correctness. 
There were two types of syntax errors in the synthesized programs, none of which was directly related to the synthesis task. First, since new types (e.g., \code{javax.security.auth.login.LoginContext}, \code{javax.security.auth. callback.CallbackHandler}, etc.) were added to the code, the corresponding \textit{import}s had to be added to the program. This is similar to when a programmer copy-pastes a code snippet from a Q\&A forum (e.g., Stack Overflow) and adds it to the code. Since the synthesizer knows the type of API calls that are used in the synthesis process, we will be able to add this feature to our synthesizer in the future. Second, there were some abstract classes or interfaces that needed customized implementations. For instance, \code{CallbackHandler} is an interface that has a single method, \code{handle(...)}, to be implemented by the programmer. Automatically implementing such methods requires a high level of communication with the programmer. Since our goal was to automatically capture the high-level specification of the tactic, we did not cover such customization in our synthesizer. However, it would be possible to augment the approach so that it takes advantage of active learning techniques for creating a low-level specification of those methods with little communication with the programmer.
%\hl{Apart from these two issues, synthesized codes were comilable}. We confirmed that by fixing the mentioned issues, compiling all the synthesized programs, and not receiving any compilation error.
%
% \subsubsection{Verification}
% \ali{To be written.}

\subsection{Evaluating the Entire Approach}
\label{sec:eval_overall}
So far, we investigated the performance of each component of our approach, \textsc{IPSynth}, separately. In this part, we run the entire approach over the test dataset and measure its performance. There are a variety of metrics (e.g., BLEU \cite{papineni2002bleu} and CodeBLEU \cite{ren2020codebleu}) developed by researchers for assessing the accuracy of code-generation approaches, however, these metrics either only focus on the syntax similarity (not necessarily the syntax correctness!) between the generated and the expected code or fail to capture some of the important semantic similarities. Therefore, in this paper, in addition to the syntax correctness of the tactic implementation, we also investigate the semantic correctness of the synthesized tactic. We compare our approach on these criteria against one of the most powerful code generators that is vastly used by programmers, ChatGPT. Additionally, we provide the run-time complexity metrics of our approach including the time and storage. 
%\ali{Different metrics for code generation evaluation, including BLEU, ROUGE-L, METEOR, ChrF, CodeBLEU, RUBY, and the -Out of the BLEU: how should we assess ...- paper}
%
\subsubsection{Semantic correctness}
%The approach supposed to implement an architectural tactic in a program. 
\textsc{IPSynth} synthesizes tactic implementations that are composed of API calls from a framework of interest. % (e.g., JAAS).
Hence, the \textit{semantic correctness} in the context of this task refers to the correctness of incorporating APIs of the framework in the program such that the tactic is implemented correctly. We assess such correctness through two different processes, \textbf{(i)} an evaluation that only focuses on \textsc{IPSynth} and its performance, and \textbf{(ii)} a comparative analysis where we compare the performance of \textsc{IPSynth} in tactic implementation tasks against ChatGPT.

\subsubsubsection{\textsc{IPSynth}-centric evaluation}
In this experiment, we study the top-ranked tactic implementation returned by \textsc{IPSynth} (i.e., top-1 returned implementation) for each of the test programs introduced in \tableref{tbl:IPSynth_eval_comparison_dataset}. Therefore, there are $20$ synthesized programs in total that were reviewed to investigate their semantic correctness w.r.t. tactic implementation.  \tableref{tbl:eval_entireapproach_semantic_manual} provides the results of this effort. In this table, the \textit{set of APIs} represents the number of APIs that were incorporated into tactic implementation tasks. Moreover, \textit{I-Control dep.} and \textit{I-Data dep.} represent control- and data-dependencies between API calls inside a synthesized piece. Finally, \textit{E-Control dep.} and \textit{E-Data dep.} represent control- and data-dependencies between API calls of the synthesized piece and their surrounding environment. Using a correct set of API calls alongside the correct establishment of I- and E-dependencies guarantees the semantic correctness of the tactic implementation. The last row (i.e., \textit{Overall}) shows the overall status of the semantic correctness of the synthesized programs, considering the results of all the other criteria in this table. As the results suggest, in $100\%$ of the cases the correct set of API calls was used for the synthesis task. Moreover, the control- and data-dependencies between API calls in the synthesized pieces (I- dependencies) were correctly established in all of the cases. However, the accuracy of establishing correct data- and control-dependencies between API calls of synthesized pieces and their environment (E- dependencies) was $85\%$. In other words, there were three programs where the synthesizer used the correct set of API, correctly created each code piece, but put them in undesired methods in the program. Our further investigation showed that due to some similarities in method naming, the synthesizer was confused and put the code pieces in unexpected methods. Although the accuracy of the approach was not $100\%$, as we will discuss in our comparative analysis, we consider this as a high-accuracy synthesis compared to the related work.   

\begin{table}
\centering
\caption{The semantic correctness inspection. I- and E- dependencies represent dependencies between API calls in a synthesized piece and dependencies between API calls and their surrounding environment, respectively. \ali{double check}}
 \label{tbl:eval_entireapproach_semantic_manual}
\begin{tabular}{P{0.25\linewidth} P{0.2\linewidth} P{0.2\linewidth} P{0.15\linewidth} } 
    \toprule
 Criteria & \# of cases & \# of correct answers & Accuracy \\ 
    \midrule
 Set of APIs & 50 & 50 & 100\% \\ 
 I-Control dep. & 50 & 50 & 100\% \\ 
 I-Data dep. & 60 & 60 & 100\% \\ 
 E-Control dep. & 40 & 34 & 85\% \\ 
 E-Data dep. & 40 & 34 & 85\% \\ 
    \midrule
  Overall Prog. & 20 & 17 & 85\%\\
    \bottomrule
\end{tabular}
\end{table}

\subsubsubsection{Comparative Analysis}
\label{sec:IPSynth_eval_comparative_analysis}
We also compared the performance of \textsc{IPSynth} against ChatGPT, a tool that programmers rely on for code generation tasks. In order to ask ChatGPT to generate the tactic code, we provided a prompt that asks ChatGPT for incorporating the JAAS framework API for adding \textit{authentication} to the given source code. Then, we shared the content of the program with ChatGPT, i.e., copy-pasted the Java file contents after the abovementioned prompt. In our comparative analysis, we considered the first response returned by ChatGPT for each prompt. All the provided prompts as well as the ChatGPT responses are available from the dataset repository \footnote{\url{https://anonymous.4open.science/r/Anonymous-82DE}}. 
%In this repository, under each program's directory, there is a directory \textit{generatedByLLM} which contains a file named \textit{ChatGPT.txt}. 
\tableref{tbl:IPSynth_eval_comparison_result} provides the results of this study. In this table, we compare the syntax correctness as well as the semantic correctness of the tactic implementation by \textsc{IPSynth} and ChatGPT. Please note that to stay fair, we only considered the first response (i.e., top-1) implementation returned by \textsc{IPSynth}. As the results show, except for one program (i.e., CharGPT failed on $P13$), both \textsc{IPSynth} and ChatGPT synthesized syntactically correct programs. The problem with the code implementation by ChatGPT for $P13$ was that it generated a code that included an invalid API (\code{subject.hasRole(...)}), i.e., a method call that does not exist in the JAAS framework. Thus, it caused a syntax error in the code generated by ChatGPT. However, when it comes to semantic correctness, ChatGPT was only able to correctly synthesize the first program (i.e., $P1$) and failed to have semantically correct implementations for the rest of the programs. For example, it generates code for part of the tactic and leaves the implementation for the rest of the tactic to the programmer. On the other side, except for three programs (i.e., $P2$, $P4$, and $P20$), \textsc{IPSynth} was able to correctly implement all the programs. The reason why \textsc{IPSynth} was not successful in its synthesis task for the two mentioned programs was that due to the existence of different methods in $P2$, $P4$, and $P20$ with similar code mapping score, \textsc{IPSynth} was confused and chose an undesired method (i.e., a method that did not align with the test label). Therefore, the synthesis result for the mentioned two programs was incorrect.

\begin{table}
\centering
\caption{Comparative study results}
 \label{tbl:IPSynth_eval_comparison_result}
     
\begin{tabular}{P{0.1\linewidth} P{0.15\linewidth} P{0.15\linewidth} P{0.15\linewidth} P{0.15\linewidth}} 
    \toprule
 \multirow{2}{0.1\linewidth}{Prog. No.} & \multicolumn{2}{c}{Syntax Correctness} & \multicolumn{2}{c}{Semantic Correctness}\\ 
  & \textsc{IPSynth} & ChatGPT & \textsc{IPSynth} & ChatGPT  \\ 

    \midrule

    \textbf{P1} & \checkmark & \checkmark & \checkmark & \checkmark\\
    \textbf{P2} & \checkmark & \checkmark & \xmark & \xmark \\
    \textbf{P3} & \checkmark & \checkmark & \checkmark & \xmark \\
    \textbf{P4} & \checkmark & \checkmark & \xmark & \xmark \\
    \textbf{P5} & \checkmark & \checkmark & \checkmark & \xmark \\
    \textbf{P6} & \checkmark & \checkmark & \checkmark & \xmark \\
    \textbf{P7} & \checkmark & \checkmark & \checkmark & \xmark \\
    \textbf{P8} & \checkmark & \checkmark & \checkmark & \xmark \\
    \textbf{P9} & \checkmark & \checkmark & \checkmark & \xmark \\
    \textbf{P0} & \checkmark & \checkmark & \checkmark & \xmark \\
    \textbf{P11} & \checkmark & \checkmark & \checkmark & \xmark \\
    \textbf{P12} & \checkmark & \checkmark & \checkmark & \xmark \\
    \textbf{P13} & \checkmark & \xmark & \checkmark & \xmark \\
    \textbf{P14} & \checkmark & \checkmark & \checkmark & \xmark \\
    \textbf{P15} & \checkmark & \checkmark & \checkmark & \xmark \\
    \textbf{P16} & \checkmark & \checkmark & \checkmark & \xmark \\
    \textbf{P17} & \checkmark & \checkmark & \checkmark & \xmark \\
    \textbf{P18} & \checkmark & \checkmark & \checkmark & \xmark \\
    \textbf{P19} & \checkmark & \checkmark & \checkmark & \xmark \\
    \textbf{P20} & \checkmark & \checkmark & \xmark & \xmark \\
    \bottomrule
\end{tabular}
\end{table}

\begin{figure}
    \centering
    \includegraphics[width=.9\textwidth]{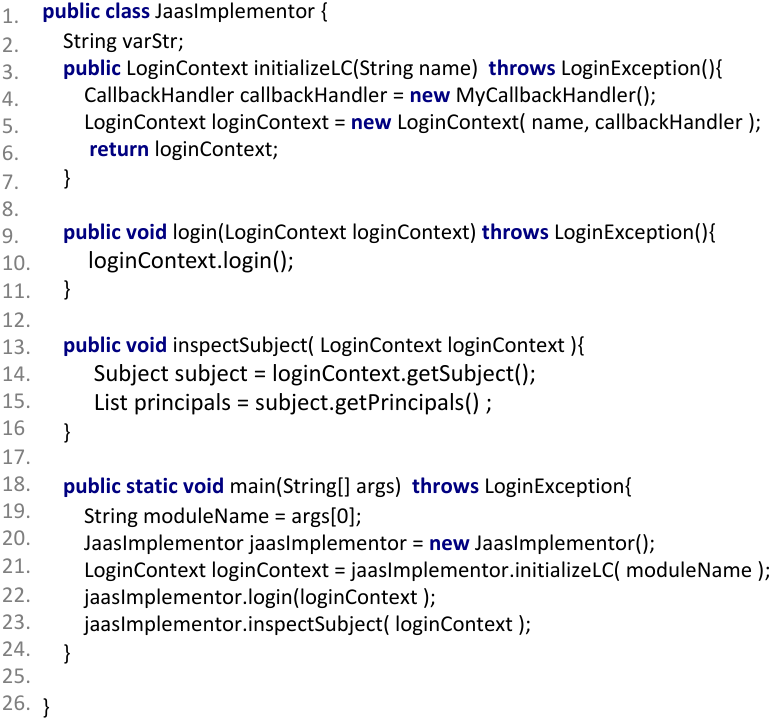}
    \caption{The expected implementation of the security tactic in the test program shown in \figref{fig:IPSynth-Dataset-TestProgram} by incorporating the JAAS framework API.}
    \label{fig:IPSynth-Dataset-TestProgram_Synthesized}
\end{figure}

\begin{figure}
    \centering
    \includegraphics[width=.6\linewidth]{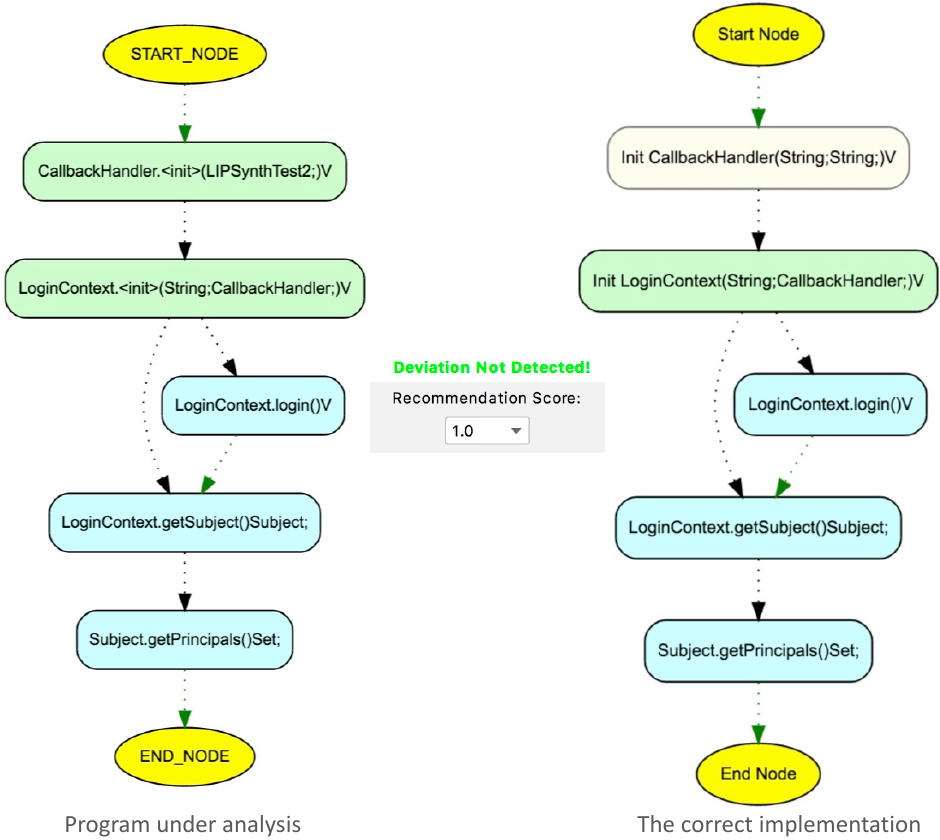}
        \centering
    \caption{A snapshot of evaluating semantic correctness of the synthesized program shown in \figref{fig:IPSynth-Dataset-TestProgram_Synthesized} by \textsc{ArCode} tool.}
    \label{fig:eval_entire_arcode_score}
\end{figure}

\subsubsubsection{Tool-based correctness demonstration}
%To avoid biasing the evaluation results based on manual inspection efforts, 
In addition to the studies on the synthesized tactics that we presented so far, we used a tool that is specialized in detecting deviations from the correct implementation of tactics in a program to demonstrate the semantic correctness of the synthesized tactics by \textsc{IPSynth}. To that extent, we used the \textsc{ArCode Plugin} tool \cite{shokri2021_icpc}, which is able to run an inter-procedural context-sensitive analysis over a program under query, create a graph-based representation of how the tactic is implemented in that program, and finally, identify possible deviations from the correct tactic implementation. Based on the performed analysis, this tool reports a score in the range of [0, 1]. The closer the score is to 0, the more deviated the program is from the correct tactic implementation. We ran this tool over the synthesized programs. The average \textsc{ArCode} score for the tactics synthesized by our synthesizer was $0.95$\ali{double check}, meaning that the tool identifies the synthesized tactics as highly correctly implemented. Recall from the \textit{comparative analysis} as well as the \textit{\textsc{IPSynth}-centric evaluation} that there were $2$ programs (among the total $20$ programs) where \textsc{IPSynth} was not able to completely implement the tactic in the expected way. However, even in those two programs, apart from a few external dependencies, the other parts of the implementation was correct. The $0.95$ score returned by the tool reflects on this phenomenon.  \figref{fig:eval_entire_arcode_score} demonstrates the result of running the \textsc{ArCode} tool on the synthesized program shown in \figref{fig:IPSynth-Dataset-TestProgram_Synthesized}. While the graph on the left side of this figure represents how API calls were used in the synthesized program, the graph on the right side visualizes the correct way of the tactic implementation, which is similar to how the synthesized program has implemented the tactic. The score in this figure measures the similarity of the implemented tactic in the synthesized program to the correct implementation. In the case of the synthesized program, the score is 1, i.e., it is identified as a correct implementation.

\subsubsection{Run-time complexities}
Finally, we measured the performance of our synthesizer in terms of the complexities of time and memory during the synthesis task. We instrumented the synthesizer so that we could capture needed statistics from each component, as well as from the entire approach. \tableref{tbl:eval_overall_complexities} presents the measured timing and memory usages per component, as well as the entire approach. We did not include FSpec annotation as it is just a one-time process in which the approach creates the required annotations. Then, unless the FSpec is changed, there would be no more FSpec annotations while synthesizing tactics. Please note that since there are some overheads when running the entire approach, the timing and memory usage presented in the last row is higher compared to the summation of corresponding columns.

\begin{table}
\centering
\caption{Run-time complexities of the synthesizer}
 \label{tbl:eval_overall_complexities}
\begin{tabular}{P{0.35\linewidth} P{0.25\linewidth} P{0.25\linewidth} } 
    \toprule
 Component & Avg. Time (ms) &  Avg. Memory (MB) \\ 
    \midrule
    Code Annotation & 155.3 & 105  \\ 
    Sketch Generation & 70.5 & 93  \\ 
    Sketch Resolving & 33.2 & 25  \\ 
    %Verification & \hl{5} & \hl{3}  \\ 
    \midrule
    Overall & 285.2 & 901  \\ 
  \bottomrule
\end{tabular}
\end{table}

\section{Limitations and Future Work}
\label{sec:Limitation}
\ali{Re-write this section based on internal and external-threats to validity}

Although the results show the effectiveness of the introduced approach in this paper, there are some limitations that need to be considered for the sake of generalizability. First, in this paper, we synthesized loop-free code snippets. We leave the task of synthesizing more complex implementation of architectural tactics to future work.
Second, our experimental study has been conducted on a popular Java security framework, JAAS. Exploring more architectural tactic enabler frameworks would be beneficial for evaluating the performance of this approach in other scenarios. Next, in this paper, we created a repository of $20$ test projects of architectural tactic synthesis tasks with different levels of complexity. Although this data set is introduced to the community for the first time, it would be interesting to investigate the performance of the approach through a wider range of tactic synthesis test cases. 
Finally, as part of the approach, we rely on method name similarities for identifying correct locations in the program for synthesizing and adding code snippets. However, in the case of obfuscated codes, this would become problematic as the method names are changed to meaningless names. To mitigate this problem, it would be possible to collect more semantic-related information around APIs while creating the FSpec. This could range from code structures to surrounding data- and control dependencies around each API call. This information could be used while computing the score of different locations in the program (\secref{sec:IPPS_CodeAnnotation}) with a higher coefficient.

\section{Related Work}
\label{sec:relatedWork}
Synthesizing a program based on a given list of components has been studied for over a decade \cite{lustig2009synthesis, jha2010oracle}. These components can range from simple instructions supported by the underlying programming language (e.g., mathematical and logical instructions) \cite{taly2011synthesizing, demarco2014automatic, zhang2020probabilistic}, to more complex components that implement crucial functionalities and use-cases (i.e., APIs) \cite{albarghouthi2013recursive, bavishi2019autopandas, iannopollo2019constrained, ellis2019write, liangcomponent, liu2020boosting}. 
\subsection{API-based Program Synthesis}
In recent years, a number of API-based program synthesis approaches have been developed. Morpheus \cite{feng2017component2} takes I/O examples of the desired to-be-synthesized R program, in addition to an optional set of components for the data manipulation task. It also incorporates a specification expressed by the user in the format of first-order logic. There have been some efforts \cite{gascon2017look} to leverage proof-based approaches to find a program that is constructed from a set of components and yet satisfies a series of constraints without exhaustively searching the program space. 
In another work \cite{tiwari2015program}, researchers apply constraints on a given list of components from functional and non-functional requirements points of view to synthesize a desired block of code. 
DAPIP \cite{bhupatiraju2017deep} learns from a repository of examples of how to concatenate APIs in order to perform specific data manipulation tasks. SyPet \cite{feng2017component} synthesizes a block of code based on a given list of APIs, the signature (i.e., type of input and output) of the expected code snippet, as well as some test cases. EdSynth \cite{yang2018edsynth} and FrAngle \cite{shi2019frangel} are able to construct a short code snippet with more control structures compared to SyPet. TYGAR \cite{guo2019program} aims to improve the performance and scalability of previous API-based synthesizers. ALPS \cite{si2018syntax} leverages a syntax-guided approach to learn from examples of Datalog programs on how to generate a sequence of Datalog rules for a given task. Hoogle${+}$ \cite{james2020digging} creates a code snippet from Haskell popular libraries based on the specified signature of the function and a set of input/output examples. In another work, SyRust \cite{takashima2021syrust} relies on the Rust language ownership type system to create a program from a sequence of Rust language API calls. RbSyn \cite{guria2021rbsyn} aims to consider the side effects of API calls as guidance while synthesizing Ruby programs. It relies on the specified type of output of the program, as well as a set of test cases. ReSyn \cite{knoth2019resource} leverages a typing system alongside a resource bound provided by the user to construct a program that s resource-efficient.
\subsection{Leveraging API Usage Knowledge}
Although the mentioned approaches use APIs as building blocks, they do not use some precious information about APIs, i.e., API usage patterns, in the synthesis process. In some of the follow-up works, researchers incorporated API usage knowledge in the synthesis task which resulted in improvement of the performance and accuracy of the synthesizer. In that regard, researchers observed a significant improvement in the run-time of state-of-the-art API-based program synthesizers (e.g., SyPet) when they benefit from pre-learned API usage patterns \cite{liu2019accelerating}. ProSy \cite{liu2020prosy} learns and uses a probabilistic API usage model in the synthesis task to reduce the program search space. This results in a reduction of the synthesis process by up to 80\% compared to SyPet. In another similar work \cite{liu2020much}, a general API usage search engine has been developed to guide the synthesis process which resulted in an 86\% reduction of synthesis time compared to SyPet. ITAS \cite{liu2020boosting2} also learns API usage models from a search engine and performs a bi-directional search strategy for improving the synthesis time. 

Despite the capabilities of all these approaches, they suffer from two shortcomings that make them incapable of being used in architectural tactic synthesis tasks. First, the need for providing specifications of the to-be-synthesized code, and second, the lack of ability to synthesize discrete (and yet related) code snippets in different locations of the program. In this paper, we introduce an approach that addresses these concerns.

\section{Conclusion}
\label{sec:Conclusion}
In this paper, we introduced \textsc{IPSynth}, a novel program synthesis approach for implementing architectural tactics in an existing code base. This approach is able to construct the to-be-synthesized code as smaller pieces of sketches, locate proper spots in the program for each piece, generate sketches corresponding to each piece, resolve all the sketches, and implement the desired architectural tactic in the program. This is an inter-procedural program synthesis approach in which the specifications are automatically collected from a pre-learned and annotated framework API specification model, as well as the structure of the program. The experimental study shows the high accuracy of the approach for tactic synthesis tasks. Moreover, the comparative analysis against related work demonstrates the effectiveness of the approach in inter-procedural code synthesis compared to the state-of-the-art.     

This approach can enable many software development tools (e.g., IDEs) to support programmers with automation of architectural tactic recommendation and implementation. Moreover, the realization of this approach as an educational tool can serve educational purposes such as teaching software architecture to program developers. 


%\bibliographystyle{IEEEtran}


% \end{thebibliography}

\bibliographystyle{IEEEtran}
\bibliography{./bibliography}

% % biography section
% % 
% % If you have an EPS/PDF photo (graphicx package needed) extra braces are
% % needed around the contents of the optional argument to biography to prevent
% % the LaTeX parser from getting confused when it sees the complicated
% % \includegraphics command within an optional argument. (You could create
% % your own custom macro containing the \includegraphics command to make things
% % simpler here.)
% %\begin{IEEEbiography}[{\includegraphics[width=1in,height=1.25in,clip,keepaspectratio]{mshell}}]{Michael Shell}
% % or if you just want to reserve a space for a photo:

% \end{IEEEbiography}


% % insert where needed to balance the two columns on the last page with
% %\newpage



%\vfill


\end{document}